\documentstyle[12pt,epsf]{article}
\def\gs{> \kern -12pt \lower 5pt \hbox{$\displaystyle{\sim}$}}
\def\ls{< \kern -12pt \lower 5pt \hbox{$\displaystyle{\sim}$}}

\def\ve{\varepsilon} 
\def\be{\begin{equation}}
\def\bea{\begin{eqnarray}}
\def\ena{\end{eqnarray}}
\def\en{\end{equation}}
\newcommand{\bi}[1]{\mbox{\boldmath$#1$}}
\newcommand{\av}[1]{\langle{#1}\rangle}
\newcommand{\tensor}[1]{\stackrel{\leftrightarrow}{{#1}}}

\def\no{\noindent}
\def\p{\partial}

\begin{document}
{\title{\Large \bf{ELECTRIC FIELD EFFECTS  NEAR CRITICAL POINTS}
\footnote{NATO series {\it Nonlinear Dielectric Phenomena 
in Complex Liquids} (Kluwer, 2003)}}

\author{AKIRA ONUKI \\
{\it Department of Physics, Kyoto University, Kyoto 606-8502, Japan}} 
\date{ }
%\begin{document}
\maketitle

\no 
{\bf Abstract}.  We present a general Ginzburg-Landau 
theory of electrostatic interactions 
and electric field effects 
for the order parameter, 
 the polarization,  and 
the charge density.  Electric field effects are then 
investigated in near-critical fluids  and liquid crystals. 
Some  new predictions are given on   
the liquid-liquid phase transition 
in  polar binary mixtures with  a small fraction of ions 
and the deformation of the liquid crystal 
order  around a charged particle.

\section{Introduction}

There can be a variety of electric field effects  
at phase transitions.  
For near-critical fluids a number of  such effects 
have  been investigated 
$[1$-$19]$. 
Examples are  weak critical 
singularity in the macroscopic dielectric constant 
  and the  refractive index, 
nonlinear dielectric effects, 
and electric birefringence. In liquid crystals 
electric field is coupled  to 
the nematic orientation \cite{PGbook}, 
but similar effects exist near the isotropic-nematic transition 
\cite{liq1}. 
For near-critical fluids without charges 
we  hereafter summarize previous results in this section.  
For  other systems such experments are not 
abundant, so we will try to predict some effects   
in the following sections.

In near-critical fluids 
the order parameter $\psi $ 
has been taken  to be   $n/n_{\rm c}-1$ for pure (one-component) 
 fluids and  to $c-c_{\rm c}$ for binary fluid mixtures, 
where $n$ is the density and $c$ is the 
molar concentration or volume fraction of one component. 
The quantities with the subscript c will 
represent the critical values.   To be precise, however, 
there is a mapping relationship between fluids and 
Ising systems in describing 
the critical phenomena \cite{Onukibook}. The number density  
 in fluids may be expressed as 
\be 
n({\bi r})  = n_0+ \alpha_1 \psi ({\bi r}) 
+ \beta_1 \psi({\bi r})^2, 
\label{eq:1.1}
\en  
in terms of the spin   $\psi$$(=$the true order 
parameter) in the corresponding 
Ising system.   Although not well justified, 
we expand $\ve$ in powers of $\psi$  in a local form as 
\be
\ve ({\bi r}) = \ve_{0} + \ve_1\psi({\bi r}) + 
\frac{1}{2}\ve_2\psi({\bi r})^2+ \cdots ,
\label{eq:1.2}
\en 
where   $\ve_1$ and $\ve_2$ are constants 
independent of  the reduced temperature 
$\tau=T/T_{\rm c}-1$.   
The quantity $\psi^2$ corresponds to 
the energy density in Ising  systems and 
the average $\av{\psi^2}$ 
over the thermal fluctuations 
contains a term proportional to 
$|\tau|^{1-\alpha}$ 
on the critical isochore above $T_{\rm c}$ 
and  on the coexistence curve below  $T_{\rm c}$ \cite{Onukibook}. 
Therefore, the weak singularity  
($\propto |\tau|^{1-\alpha}$) 
appears in the so-called  coexistence curve 
diameter   for $\beta_1\neq 0$ below $T_{\rm c}$ 
and in the thermal average $\av{\ve}$ 
for $\ve_2 \neq 0$ above $T_{\rm c}$.

Regarding the expansion (1.2) we 
make two remarks.  (i) For nonpolar pure fluids  
the overall density-dependence of  
$\ve$ may well be approximated by 
the Clausius-Mossotti relation  
$({\ve-1})/({\ve+2})=  \alpha_{\rm m} n/3$,  
where $\alpha_{\rm m}$ is the molecular polarizability.  
Using this  relation and assuming 
$\beta_1=0$ in (1.1) we have   
$\ve_1 =  (\ve_0-1)(\ve_0+2)/3$ 
and $\ve_2 = 2\ve_1^2/(2+\ve_0)>0$.
However, dielectric formulas like 
 Clausius-Mossotti are valid only in the long wavelength 
limit   where the electric field 
fluctuations are   neglected (in text books) 
or averaged out \cite{Mazur,Stell}. Thus  estimation of 
 $\ve_2$ from Clausius-Mossotti 
is not justified when  we use 
(1.2) to derive dielectric anomaly 
due to   the small-scale critical 
fluctuations.   Notice that (1.1) 
itself gives rise to 
a contribution $(=2\beta_1\ve_1/\alpha_1)$ to 
$\ve_2$ from the linear density 
dependence $\ve=\ve_{\rm c} 
+ \ve_1(n/n_{\rm c}-1)$. (ii) For binary mixtures A+B, $\ve_1$ 
should be of order 
$\ve_{\rm A}-\ve_{\rm B}$, where  
$\ve_{\rm A}$ and $\ve_{\rm B}$ are 
the dielectric constants of the two components, so 
$\ve_1$  can be   10-100 in polar binary mixtures 
(where  at least one component is  polar). 
Electric field effects are  much stronger  in  
polar binary mixtures than in nonpolar fluids.

Once we assume  (1.2), 
 it is straightforward to solve 
the Maxwell equations 
within the fluid  if the  deviation 
$\delta \ve= \ve - \av{\ve}$ is treated as a small 
perturbation.   
Hereafter $\av{\cdots}$ represents 
the average over the  fluctuations of $\psi$. 
The macroscopic static  dielectric tensor 
$\tensor{\ve}_{\rm eff}$ is obtained from 
 the relation $\av{{\bi D}}= \tensor{\ve}_{\rm eff}\cdot\av{{\bi E}}$ 
between  the average electric induction $\av{{\bi D}}$ 
and the average electric field $\av{{\bi E}}$.   
To leading order, 
the fluctuation contribution $\Delta{\tensor \ve}_{\rm eff}$  
  in the long wavelength limit 
can be expressed in terms of 
the structure factor  $I({\bi q})= 
\av{|\psi_{\small{\bi q}}|^2}$ \cite{Mazur,Onuki_bi1}.  The   
contribution  from the long  wavelength fluctuations 
 with $q<\Lambda$ is   written as 
\be
\Delta{\tensor \ve}_{\rm eff}^<   =  
\int_{\bi q} I({\bi q})  
\bigg   ( \frac{1}{2}\ve_2 \tensor{I} 
- \frac{1}{\ve_0}\ve_1^2   {\hat{\bi q}\hat{\bi q}} \bigg ) , 
\label{eq:1.3}
\en 
where $\int_{\bi q}= (2\pi)^{-3}\int d{\bi q}$, 
$\tensor{I}=\{\delta_{ij}\}$ is the unit tensor,   and 
$\hat{\bi q}= q^{-1}{\bi q}$. The 
upper  cutoff wave number $\Lambda$ in (1.3) 
is smaller than the inverse particle size. 
Note that $I({\bi q})$ is uniaxial at small $q$   
in the direction of electric field, but it is nearly 
isotropic at large $q$. 
Thus the contribution $\Delta{\tensor  \ve}_{\rm eff}^>$ 
from the short wavelength fluctuations  with $q>\Lambda$ is 
proportional to the unit tensor  
and  may be  included into $\epsilon_0$ in (1.2).

We may also calculate an 
electromagnetic wave 
within near-critical fluids \cite{Mazur,Onuki_bi1}. 
Let $\bi k$ be its wave vector.  
The average electric field $\av{{\bi E}}$ 
is nearly perpendicular to $\bi k$, oscillates with 
frequency $\omega$,  and obeys 
\be 
\omega^2 \bigg [ \av{\ve (\omega)} + 
\Delta{\tensor \ve}({\bi k},\omega)  \bigg ]\cdot \av{{\bi E}}
= c^2k^2 \av{{\bi E}}, 
\label{eq:1.4}
\en 
where $\ve(\omega)=\ve_0(\omega) 
+\ve_1(\omega)\psi+\cdots$ is the dielectric 
constant at frequency $\omega (\cong ck/\ve_0(\omega)^{1/2})$  
dependent on $\psi$ 
as in (1.2). The fluctuation contribution to
 the dielectric tensor for the electromagnetic waves 
is written as 
\cite{Mazur,Onuki_bi1,Onuki_bi2}
\be 
\Delta{\tensor \ve}({\bi k},\omega)
= \frac{\ve_1(\omega)^2}{\ve_0(\omega)} 
\int_{\bi q}  \frac{I({\bi k}-{\bi q})}{q^2- k^2-i0} 
(k^2{\tensor I}_\perp- {\bi q}_\perp{\bi q}_\perp  ), 
\label{eq:1.5}
\en 
where  ${\bi q}_\perp
= {\bi q}- k^{-2} ({\bi q}\cdot{\bi k}){\bi k}$ 
and $\tensor{I}_\perp= {\tensor I}- k^{-2}{\bi k}{\bi k}$  
are  perpendicular to ${\bi k}$,  
and $+i0$ represents a small 
imaginary part arising from the causality law.
The imaginary part 
gives rise to  damping  and can be calculated using 
 the formula  
$1/(x-i0)= {\rm vp}(1/x) +i\pi \delta (x)$ with vp 
representing taking the Cauchy principal value. 
Then $\Delta{\tensor \ve}({\bi k},\omega)$ has two 
eigenvalues  $\Delta_\alpha (\omega)$ 
($\alpha=1,2$) in the plane perpendicular to $\bi k$. 
The dispersion relations for the two polarizations 
are written as  
\be 
k= c^{-1} \omega 
[ \av{\ve(\omega)}+ \Delta_\pm (\omega)]^{1/2}.
\label{eq:1.6}
\en 
In particular, if we apply an electric field along the 
$z$ axis and send a laser beam along the $x$ axis, 
we have $\Delta \ve_{yz}=
\Delta \ve_{zy}=0$. The  principal polarization  
is then either along the $z$ axis 
($\alpha=1$) or along  the $y$ axis ($\alpha=2$). 
The turbidity ${\cal T}_\alpha$ 
is written as 
\be 
{\cal T}_\alpha= \frac{k}{\ve_0(\omega)}
 {\rm Im} \Delta_\alpha (\omega) 
= \frac{\omega^4}{16\pi^2c^4} 
{\ve_1(\omega)^2} \int d\Omega I({\bi k}-k{\hat{\bi q}})
(1- \hat{q}_\alpha^2  ),  
\label{eq:1.7}
\en 
where $\hat{q}_1= \hat{q}_z$, 
  $\hat{q}_2= \hat{q}_y$, and 
 $d\Omega$ represents integration over the direction 
$\hat {\bi q}$.  The Ornstein-Zernike 
intensity gives the famous  turbidity expression.  
Generally, when $I({\bi q})$ depends on  $\hat{\bi q}$, 
the difference $\Delta_1-\Delta_2$  
becomes nonvanishing. Its real and imaginary parts 
can be measured as  (form) 
 birefringence and dichroism, respectively. 
In near-critical fluids under shear flow,  
 both these  effects  exhibit strong 
critical divergence \cite{bifluids}.   
In polymer science, 
 birefringence in shear flow 
arising from the above origin is called 
form birefringence \cite{Onuki_bi2}, whereas 
  alignment of optically anisotropic  particles   
gives rise to  anisotropy in $\av{\ve_{ij}}$ 
leading to intrinsic birefringence.  Form 
dichroism is maximum  when the scattering objects 
have sizes of the order of the laser  wavelength 
and,  for shear flow, it has been used to detect anisotropy of 
the particle pair correlations in 
colloidal suspensions \cite{Fuller0} 
and polymer solutions \cite{Fuller_e2}. 
In Ref.{\cite{Onuki_bi1}}  an 
experimental setup to measure dichroism 
was  illustrated.

In Section 3 we will examine 
the effects mentioned above using (1.3) and (1.5).

\setcounter{equation}{0} 
\section{Ginzburg-Landau free energy in electric field}

We will construct a Ginzburg-Landau 
free energy   including  electric field   under each given 
 boundary condition, since  there seems to be 
no clear-cut argument on  its form  in the literature. 
In our scheme 
 the gross variables are the order parameter $\psi$, 
the  polarization $\bi p$, and the  charge 
density $\rho$. They are  coarse-grained 
variables with their   microscopic space  variations 
being  smoothed out. Since the electromagnetic field 
is determined for instantaneous values of 
the gross variables, we will 
suppress their  time dependence.

\subsection{Dielectrics under given capacitor charge}

\begin{figure}[t]
\vspace{-6cm}
\epsfxsize=6.0in
\centerline{\epsfbox{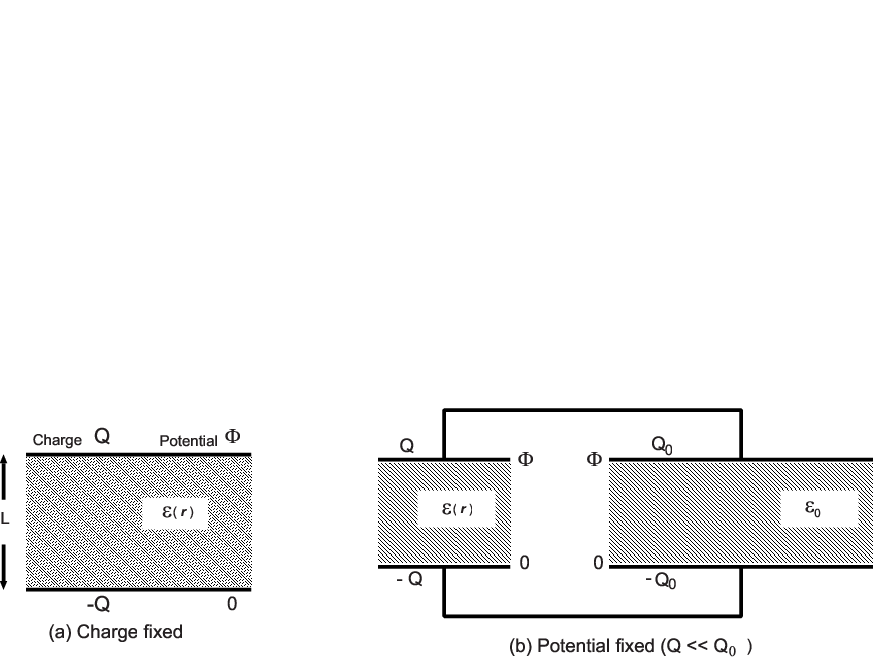}}
\caption{\protect%\narrowtext
\small{(a) System of a capacitor and a dielectric material with 
inhomogeneous dielectric constant $\ve({\bi r}) $  at 
fixed capacitor charge $Q$. The potential difference 
$\Phi$   is a fluctuating quantity dependent on 
$\ve({\bi r})$.  (b) Two 
capacitors connected in parallel with charges $Q$ and $Q_0$. 
 The smaller one  contains  
an inhomogeneous dielectric material,  
and the  larger one 
a homogeneous dielectric material. In the limit  $Q/Q_0 \rightarrow 0$, 
the potential difference $\Phi$ becomes fixed, 
while $Q$ is a fluctuating quantity. 
  }}
\label{1}
\end{figure}

The first typical experimental geometry 
is  shown in Fig.1a, where we insert our system  between 
two  parallel metallic plates 
with area $S$ and separation distance $L$. 
We assume $S^{1/2} \gg L$ and neglect the effects of 
edge fields. Generalization to other 
geometries is straightforward. 
The $z$ axis is taken perpendicularly  to the plates.  
Let the average  
surface charge density of the  upper plate be 
$\sigma_{\rm ex}$ and that of the  
lower plate  be 
$-\sigma_{\rm ex}$. 
The total charge on the upper plate is 
\be 
  Q= S\sigma_{\rm ex}. 
\label{eq:2.1}
\en 
The  Ginzburg-Landau free energy functional 
$F$ consists of a  chemical   
part  $F_{\rm ch}\{\psi,{\bi p}, \rho\}$ and an 
 electrostatic part 
$F_{\rm st}\{{\bi p},\rho, Q\}$ as
\be 
F=F_{\rm ch}\{\psi,{\bi p},\rho \}+F_{\rm st}\{{\bi p}, \rho, Q\}, 
\label{eq:2.2}
\en 
where  $\psi$   represents 
a set of  variables including  the order parameter. 
In ferroelectric systems 
  ${\bi p}$ is the order parameter. 
The equilibrium 
distribution of the gross variables   is given by 
const.$\exp (-F/k_{\rm B}T)$ at each fixed $Q$. 
We determine $F_{\rm st}$  as follows. 
If  infinitesimal deviations $\delta {\bi p}$, $\delta\rho$, 
and $\delta Q$ 
are superimposed on ${\bi p}$, $\rho$,  and $Q$, 
the incremental change of  
$F_{\rm st}$ should be  given by 
the work done by the electric field, 
\bea 
\delta F_{\rm st}&=&F_{\rm st}\{{\bi p}+ \delta{\bi p},\rho+\delta\rho, Q+ 
\delta Q\}-  F_{\rm st}\{{\bi p}, \rho,Q \}\nonumber \\ 
&=& \int d{\bi r}[-{\bi E}\cdot{\delta {\bi p}}+\phi\delta\rho] 
  + \Phi \delta Q, 
\label{eq:2.3}
\ena 
where the space integral is within 
the system between the plates, 
${\bi E}= -\nabla\phi$ is the electric field vector, 
and $\Phi$ is the potential 
difference between the two plates.   
The electric potential $\phi$ 
may be set equal to $0$ at the lower plate and 
$\Phi$ at the upper plate. 
The  electric induction $\bi D= {\bi E}+ 4\pi {\bi P}$ 
satisfies 
\be 
\nabla \cdot {\bi D}= -\nabla^2\phi + 4\pi \nabla\cdot{\bi p}=  4\pi \rho,  
\label{eq:2.4}
\en 
in the bulk region. The  potential $\phi$ satisfies 
\be 
-\nabla^2\phi=4\pi\rho_{\rm eff},
\label{eq:2.5}\en 
where 
\be 
\rho_{\rm eff}({\bi r}) = \rho({\bi r})  - 
\nabla\cdot {\bi p}({\bi r})   
\label{eq:2.6}
\en 
is the effective charge density.  
The boundary conditions at $z=0$ and $L$ are  
\be 
E_x=E_y=0, \quad D_z= -4\pi\sigma_{\rm ex} . 
\label{eq:2.7}
\en  
With these relations of electrostatics 
we can integrate (2.3) formally  as 
\be 
F_{\rm st}= \frac{1}{8\pi} \int d{\bi r} {\bi E}^2. 
\label{eq:2.8}
\en 
In fact (2.8) leads to the second line of (2.3) 
 if use is made of  $\delta ({\bi E}^2)= 
-2{\nabla \phi}\cdot\delta{\bi D}-8\pi{\bi E}\cdot\delta{\bi p}= 
-2\nabla (\phi\delta{\bi D})+8\pi(\phi \delta\rho-{\bi E}\cdot\delta{\bi p})$.

To explicitly express 
$F_{\rm st}$ in terms of the gross variables, 
we first assume that 
all the quantities depend only on $z$ and ${\bi p}$ 
is along the $z$ axis, for simplicity 
\cite{Binder}. 
If we define $w(z)= \int_0^z dz' \rho(z')$, 
we obtain 
\be 
\phi (z)= 4\pi \int_0^z dz' [p_z(z')+ \sigma_{\rm ex} -w(z')]. 
\label{eq:2.9}
\en 
From the overall charge neutrality  condition 
we  require $w(L)=0$, so (2.7) is satisfied and  
\be 
F_{\rm st}= \frac{S}{8\pi} \int_0^L dz E_z(z)^2 = 2\pi S \int_0^L  dz 
[p_z (z) + \sigma_{\rm ex} -w(z)]^2  . 
\label{eq:2.10}
\en

For  general inhomogeneous $\bi p$ and $\rho$, we 
define the lateral averages, 
\be 
\bar{\bi p}(z) = \frac{1}{S} \int d{\bi r}_\perp  {\bi p}({\bi r}), \quad 
{\bar\rho}(z) =  \frac{1}{S} \int d{\bi r}_\perp 
\rho ({\bi r}), 
\label{eq:2.11}
\en 
where ${\bi r}_\perp= (x,y)$ is the lateral 
position vector. We may assume  
$\bar{p}_x={\bar p}_y=0$ from the geometrical symmetry 
in the limit $S^{1/2}/L \rightarrow \infty$. 
The effective charge density $\rho_{\rm eff}$ is 
divided into the lateral average 
$\bar{\rho} - d{\bar{p}_z}/dz$ and 
the inhomogeneous part, 
\be 
\rho_{\rm inh}({\bi r}) = [\rho({\bi r})  -{\bar\rho}(z)]-  
\nabla\cdot [{\bi p}({\bi r})-{\bar{\bi p}}(z) ].   
\label{eq:2.12}
\en   
The  electric potential is  expressed as 
\be
\phi({\bi r}) =  
4\pi\int_0^z dz'  [{\bar p}_z(z')+ \sigma_{\rm ex} -{\bar w}(z')] 
+ 4\pi\int d{\bi r}'G({\bi r}, {\bi r}') \rho_{\rm inh} ({\bi r}')  . 
\label{eq:2.13}
\en   
The first term is of the same form as (2.9) and 
\be 
{\bar w}(z) = \int_0^z dz'  {\bar\rho} (z'),
\label{eq:2.14}
\en 
which vanishes at $z=0$ and $L$. 
The  Green function 
$G({\bi r}, {\bi r}')= G({\bi r}', {\bi r})$ satisfies 
\be 
\nabla^2G({\bi r}, {\bi r}')=-  \delta({\bi r}- {\bi r}')
\label{eq:2.15}
\en  
and vanishes as ${\bi r}$ and ${\bi r}'$ 
approach the surfaces of the conductors as 
\be 
 \lim_{z\rightarrow 0,L}G({\bi r}, {\bi r}')= 
\lim_{z'\rightarrow 0,L}G({\bi r}, {\bi r}')=0.
\label{eq:2.16}
\en  
Then we have $E_x=E_y=$ at $z=0$ and $L$. 
The potential difference $\Phi$ is written as 
\bea 
\Phi&=&  
4\pi\int_0^L dz  [{\bar p}_z(z)+ \sigma_{\rm ex} -{\bar w}(z)] \nonumber\\
&=& 4\pi L \av{p_z+\sigma_{\rm ex}  +z \rho}_{\rm s}, 
\label{eq:2.17}
\ena
where $\av{\cdots}_{\rm s}= \int d{\bi r}(\cdots)/SL$ 
represents  the space average 
and use has been made of $\av{\rho}_{\rm s}=0$.  This relation 
may also be written as 
\be 
\Phi/4\pi L-  \av{p_z}_{\rm s}=  Q/S   -\av{(L/2 -z) \rho}_{\rm s},
\label{eq:2.18}
\en 
which  relates  the mean  electric field 
$-\Phi/L$   and the capacitor charge $Q$. 
The second  term on the right hand side becomes  important 
when the charge is accumulated near the capacitor plates in 
electric field.

Because we are taking 
 the limit $S^{1/2}/L \rightarrow \infty$,  
 the translational invariance $G({\bi r},{\bi r}')= 
G({\bi r}_\perp-{\bi r}_\perp',z,z')$ holds in the $xy$ plane. 
The Fourier transformation 
$G_k(z,z')$
%= 
%\int d{\bi r}_\perp G({\bi r},{\bi r}')
%\exp [i{\bi k}\cdot({\bi r}_\perp-{\bi r}_\perp')]$ 
in the 
$xy$ plane  satisfies $(-\nabla_z^2+k^2) G_k(z,z')= 
 \delta(z-z')$ and  becomes \cite{Onuki-Doi4}
\bea 
G_k(z,z')&=& 
G_k(z',z)=\frac{1}{2k} e^{-k|z-z'|}-\frac{1}{2k\sinh (kL)} 
 \nonumber\\ 
\vspace{0.5mm}
&&\hspace{-2cm} \times 
\bigg [ \sinh (kz)e^{-k(L-z')} +   \sinh (kL-kz)e^{-kz'}  
\bigg ]. 
\label{eq:2.19}
\ena
In particular, in the long wavelength limit $k \rightarrow 0$, we have 
\be 
G_0(z,z')=
\lim_{k\rightarrow 0} G_k(z,z')=\frac{1}{2}
 [ z+z'-|z-z'|] -  zz'/L   . 
\label{eq:2.20}
\en 
From  (2.17) and (2.20), as it should  be the case, 
 the  potential $\phi$ 
is also  expressed as 
\be 
\phi({\bi r})= \Phi z/L  + 4\pi 
\int d{\bi r}'G({\bi r}, {\bi r}') 
\rho_{\rm eff} ({\bi r}')  .
\label{eq:2.21}
\en 
The electrostatic energy is written 
in terms of $\rho_{\rm inh}$ as    
\be 
F_{\rm st}= 2\pi S \int_0^L  dz \bigg 
[{\bar p}_z (z) + \sigma_{\rm ex} -w(z)\bigg ]^2 
+2\pi \int d{\bi r}\int d{\bi r}'\rho_{\rm inh}({\bi r}) 
G({\bi r},{\bi r}')  \rho_{\rm inh}({\bi r}') . 
\label{eq:2.22}
\en 
With the aid of  (2.17) and (2.20),  $F_{\rm st}$ 
in terms of  $\rho_{\rm eff}$  reads  
\be 
F_{\rm st}= \frac{S}{8\pi L}\Phi^2 +   
2\pi \int d{\bi r}\int d{\bi r}'\rho_{\rm eff}({\bi r}) G({\bi r},{\bi r}')  \rho_{\rm eff}({\bi r}') .
\label{eq:2.23}
\en

\subsection{Dielectrics under given capacitor potential}

In the previous case, the capacitor  charge $Q$ is a control 
parameter and the potential difference $\Phi$ is a fluctuating 
quantity. We may  also control  $\Phi$ 
 by using  (i) a battery at a fixed potential difference 
connected to the capacitor or (ii) another large 
capacitor connected in parallel to 
the capacitor containing a dielectric material   
as  in Fig.1b.    We examine the well-defined case (ii).
The  area $S_0$ and the charge $Q_0$ of the large capacitor 
are  much larger  than 
$S$ and $Q$, respectively, so 
the large capacitor acts as a charge reservoir. 
We are supposing  an experiment in which 
 the total charge $Q_{\rm tot}=Q_0+Q$ is fixed and the 
potential difference  
is commonly given by 
$\Phi=  Q_0/C_0$, where $C_0$ is the capacitance 
of the large capacitor. Obviously, in 
the limit   $Q/Q_0 \sim S/S_0 \rightarrow 0$, 
the deviation of $\Phi$ from the   upper bound  
$\Phi_{\rm tot}=Q_{\rm tot}/C_0$ becomes  
negligible.  Because the  electrostatic energy 
of the large capacitor is given by 
$E_0 = Q_0^2/2C_0= (Q_{\rm tot}^2/2C_0)  
(1-Q/Q_{\rm tot})^2$, we obtain   
\be 
E_0 \cong   (Q_{\rm tot}^2/2C_0)-\Phi Q, 
\label{eq:2.24}
 \en 
where the first term is constant 
and a term of order $(Q/Q_{\rm tot})^2$ is neglected.     
Therefore, for  the total system  including 
the two capacitors, the potential difference $\Phi$ 
is a control parameter  and 
the appropriate free energy functional is given by  
the Legendre transformation \cite{Landau4}, 
\be 
G= F - \Phi Q . 
\label{eq:2.25}
\en   
 If $\Phi$ is fixed, 
$Q$ is a fluctuating quantity  determined by (2.18). 
The equilibrium distribution of the gross variables 
is given by const.$\exp(-G/k_{\rm B}T)$ at each 
fixed $\Phi$. 

\subsection{Dielectrics  in vacuum  and dipolar ferromagnets }

There is another typical case, in which 
a sample of dielectrics is placed in vacuum. 
There is no polarization and charge immediately  
outside the sample, but there can be 
an  applied electric field 
 created by  capacitor plates 
  far from the sample.  As far as I am aware, 
simulations of phase 
transitions in  dipolar particle systems  have   been performed 
under this boundary condition \cite{Groh}, where  the particles 
are assumed to interact via the dipolar potential 
$\propto \delta_{ij}/r^3-3x_ix_j/r^5$ 
even close to the surface. In  
ferromagnetic systems there is 
no magnetic charge 
and this boundary condition is always assumed \cite{dipolar1}.

 We  assume 
no free charge ($\rho=0$) and apply  
an electric field $E_{\rm ex}$ 
along the $z$ axis. The  electrostatic potential  
 is given by 
\bea 
\phi({\bi r})&=& -E_{\rm ex} z + \int d{\bi r}' G_0({\bi r},{\bi r}') 
\rho_{\rm eff}({\bi r}') + 
\int da G_0({\bi r},{\bi r}_a) 
\sigma_{\rm eff}({\bi r}_a)
 \nonumber\\
&=& -E_{\rm ex} z + \int d{\bi r}'{\bi p}({\bi r}')\cdot
\nabla'G_0 ({\bi r}-{\bi r}').  
\label{eq:2.26}
\ena 
In the first line 
$\rho_{\rm eff}= -\nabla\cdot{\bi p}$ 
is the effective charge 
as given in  (2.6),  $\sigma_{\rm eff}= 
{\bi p}\cdot{\bi n}$ is the effective surface charge 
with $\bi n$ being the outward 
normal  unit vector at the surface,  and 
${\bi r}_a$ represents the surface point. 
In the second line the space 
integral  is within  the sample  and 
$\nabla'= \p/\p {\bi r}'$.  
Since there is no 
capacitor charge in contact with 
the system, $G_0$ is of the usual Coulombic form, 
\be 
G_0({\bi r}-{\bi r}') = {1}/{|{\bi r}-{\bi r}'|} . 
%\frac{1}{|{\bi r}-{\bi r}'|} . 
\label{eq:2.27}
\en 
The total free energy consists of the chemical part and 
electrostatic part as 
$F= F_{\rm ch}\{\psi,{\bi p}\} 
+ F_{\rm st}\{{\bi p}\}$.    The latter  is the usual dipolar 
interaction, 
\be
F_{\rm st} = -\int d{\bi r}  E_{\rm ex}p_z 
+ \frac{1}{2}\int d{\bi r} \int d{\bi r}'   
\sum_{i j}p_i({\bi r})p_j({\bi r}')
[-\nabla_i\nabla_j G_0({\bi r}-{\bi r}')], 
\label{eq:2.28}
\en  
which satisfies  $\delta  F_{\rm st}/\delta {\bi p}= 
-{\bi E}$  in  (2.3).  Here 
   $-\nabla_i\nabla_j G_0({\bi r}) 
=  \delta_{ij}/r^3-3x_ix_j/r^5$, as ought to be the case. 
In the ferromagnetic case,  
 ${\bi p}$ and ${E}_{\rm ex}$ correspond 
to  the magnetization 
and the applied magnetic field, respectively \cite{Landau4}.

The above electrostatic (magnetic) 
energy (2.28) depends on the sample shape due to 
the depolarization (demagnetization) 
effect, so the phase transition to a ferroelectric 
(ferromagnetic) phase  becomes shape-dependent \cite{Groh}. As is well known, 
if  the dielectric constant $\ve$ is  uniform 
in the sample,  the  field inside  is given by 
$E_{\rm ex}/\ve$ 
for a thin plate 
and $3E_{\rm ex}/(2+\ve)$  
for a sphere along the 
$z$ axis.

\section{Near-critical fluids without ions} 
\setcounter{equation}{0} 

As an application of the general 
scheme in Section 2, we  
consider   a near-critical fluid 
 in the geometry of Fig.1a. 
We will use the results already presented in Section 1.

\subsection{General relations in the charge-free case}

In the absence of  free charges ($\rho=0$), 
  $F_{\rm ch}$ in (2.2) is written as 
\be 
F_{\rm ch}\{\psi, {\bi p}\} 
= F_0\{\psi\}+ \int d{\bi r}\frac{1}{2\chi}{\bi p}^2  , 
\label{eq:3.1}
\en 
where $\chi$ represents  the electric susceptibility. 
The first term is  the usual free energy functional 
for the order parameter $\psi$. Close to the critical point 
it is of the form,  
\be 
F_0\{\psi\}=  k_{\rm B}T_{\rm c}  \int d{\bi r} \bigg [ 
\frac{a_0}{2}\tau \psi^2+ \frac{u_0}{4}\psi^4 +\frac{K_0}{2} 
|\nabla\psi|^2 \bigg ],
\label{eq:3.2}
\en 
where  
$a_0$, $u_0$ and $K_0$ are 
 positive  constants,    and  
$\tau$ is the reduced temperature.  
The term $\propto \int d{\bi r}\psi$ 
is   not written (because $\psi$ 
is a conserved variable).    Use of (2.3) gives 
\be 
\delta F
= \delta F_0+ 
\int d{\bi r}\bigg [ (\chi^{-1}  {\bi p}-{\bi E}) 
\cdot{\delta {\bi p}} - \frac{1}{2\chi^2} {\bi p}^2 \delta\chi 
\bigg ] 
  + \Phi \delta Q.  
\label{eq:3.3}
\en  
Thus $F$ is minimized for 
\be 
{\bi p}= \chi {\bi E}, \quad  
{\bi D}= \ve {\bi E}.
\label{eq:3.4}
\en 
The static dielectric constant is  
\be 
\ve= 1+4\pi \chi
\label{eq:3.5}
\en 
and  depends locally on 
the order parameter $\psi$  as in (1.2).  In usual fluids 
 ${\bi p}$ rapidly 
relaxes to $\chi{\bi E}$  on a microscopic time scale, so 
(3.4) may be assumed  even in 
nonequilibrium. 
Then $\bi p$ is determined by 
$\psi$ and $Q$.  
The charge-free condition within the fluid 
$\nabla\cdot{\bi D}=0$ is written as  
\be 
\nabla \cdot (\ve \nabla \phi)=0.  
\label{eq:3.6}
\en

From the general expressions (2.8) and (3.1) 
with the aid of (3.4) we obtain 
\be 
F  = F_0\{\psi\} + \frac{1}{8\pi} 
\int d{\bi r}{\bi E}\cdot{\bi D} 
= F_0\{\psi\} + \frac{1}{2} \Phi Q.
\label{eq:3.7}
\en 
The  expression in terms of $Q$ 
 also follows directly from 
integration of $\delta F= \Phi \delta Q$ at fixed $\psi$, 
where the factor  $1/2$ arises because 
the ratio $\Phi/Q$ is a functional of $\psi$ only 
and is independent of $Q$  from   (3.4). 
 On the other hand, in the fixed-potential  case 
 in the geometry of Fig.1b, 
the  free energy functional  $G$ in (2.25) is written as 
\be
G = F_0\{\psi\} - \frac{1}{8\pi} 
\int d{\bi r}{\bi E}\cdot{\bi D} = F_0\{\psi\}-\frac{1}{2} \Phi Q.  
\label{eq:3.8}
\en
The electrostatic parts   
$\Delta F=F-F_0 $  and $\Delta G=G-F_0 $ are 
 functionals of $\ve({\bi r})$ 
at fixed $Q$ and $\Phi$, 
respectively.  The  functional derivative  of 
 $\Delta F $ at fixed $Q$ and 
that of  $\Delta G $  
at fixed $\Phi$ with respect to $\ve$ are both 
of the same form,  
\be
\bigg ( \frac{\delta}{\delta \ve} {\Delta F}\bigg )_Q = 
\bigg ( \frac{\delta}{\delta \ve} {\Delta G} \bigg )_\Phi = 
-\frac{1}{8\pi}{|\bi E|}^2, 
\label{eq:3.9}
\en  
because $\int d{\bi r}
{\bi E}\cdot\delta{\bi D}=
-\int d{\bi r}\nabla{\phi}\cdot\delta{\bi D}=0$ at fixed $Q$ 
and 
 $\int d{\bi r}{\bi D}\cdot\delta{\bi E}=0$ 
at fixed $\Phi$. 
These relations directly follow from (3.3) under
 ${\bi p}=\chi{\bi E}$.  
Even if  a sample is placed in vacuum, 
  $(\delta \Delta F/\delta \ve)_{E_0}$  
assumes the same form. 
To derive  (3.9) more evidently, we may 
 assume that $\ve=\ve(z)$ is a function of $z$ only. Then  
we explicitly obtain $\Delta F= 2\pi Q^2\int_0^L dz \ve(z)^{-1}$ 
and $\Delta G= -(8\pi)^{-1}S \Phi^2/\int_0^L dz \ve(z)^{-1}$ 
to confirm  (3.9).

\subsection{Landau expansion and critical behavior}  

We assume that a constant electric field 
${\bi E}_0=(0,0,-E_0)$ is applied 
in  the negative $z$ direction. 
Then  
$\ve_2E_0^2/4\pi k_{\rm B}T_{\rm c} 
(=A_0 \tau_{\rm c}$ in (3.19)) 
 and $(\ve_1^2/\ve_0)E_0^2/4\pi k_{\rm B}T_{\rm c}(=  
g_{\rm e}$ in (3.16)) 
are two relevant parameters representing the 
influence  of electric field. We assume that 
they are  {\it independent} of 
the upper cut-off $\Lambda$ of the coarse-graining, 
while the coefficients in $F_0$ in (3.2) depend on 
$\Lambda$ \cite{Onukibook}. This causes 
some delicate issues.  Theory 
should be made such that  the observable 
quantities are  independent of 
$\Lambda$.

We expand the electrostatic free energy 
with respect to the order parameter $\psi$.
In the following we assume  the fixed charge condition, 
but the same expressions  follow  in the fixed potential 
condition owing to (3.9).  
Let the electric field be  written as 
 ${\bi E}= {\bi E}_0 - \nabla  \delta\phi$,  
where   $\delta\phi$ is the deviation  of the 
electric potential 
induced by $\delta\ve= \ve-\av{\ve}$.   
 Then (3.9) is expanded as 
\be 
\bigg ( \frac{\delta}{\delta \ve} {\Delta F}\bigg )_Q   = - 
\frac{1}{8\pi}E_0^2 +   
\frac{1}{4\pi}{\bi E}_0\cdot\nabla\delta\phi +\cdots  . 
\label{eq:3.10}
\en 
To  first order in $\delta\ve$ 
the charge-free condition (3.6)  becomes    
$
{\av{\ve}}\nabla^2\delta\phi=  
{\bi E}_0\cdot\nabla\delta\ve$. 
Using 
the Green function $G({\bi r}, {\bi r}')$ in (2.15)  
 and  $\av{\ve}\cong \ve_0$, we obtain 
\be 
 \delta\phi({\bi r})= - \frac{1}{4\pi\ve_0} 
\int d{\bi r}'G({\bi r}, {\bi r}')
{\bi E}_0\cdot\nabla' \delta\ve({\bi r}').   
\label{eq:3.11}
\en
The electrostatic free energy  
${\Delta F} = F-{F}_0$ is composed of two parts up to order $O(\delta\ve^2)$. 
The first part is 
\be
{F}_{e0}  = - \frac{1}{8\pi} \int d{\bi r}  
{E}_0^2 {\ve}= - \frac{1}{8\pi} \int d{\bi r}  
{E}_0^2 \bigg ({\ve}_{0}+ \ve_1 \psi + 
\frac{1}{2}\ve_2\psi^2 \bigg ) . 
\label{eq:3.12}
\en 
The linear term $(\propto \int d{\bi r} \psi)$ 
is irrelevant for fluids and 
the bilinear term $(\propto \int d{\bi r} \psi^2)$ 
gives rise to a  shift of the critical temperature 
as will be shown in (3.19) and  (3.22). 
The second part is a dipolar  interaction 
arising  from 
the second term in (3.10),  
\be
{F}_{\rm dip} =  \frac{1}{8\pi}A_{\rm s} E_0^2  
\int d{\bi r} \int d{\bi r}' 
[ \nabla_z\psi({\bi r})] 
 G({\bi r}, {\bi r}') 
 [ \nabla_z\psi({\bi r}')  ],  
\label{eq:3.13}
\en 
where $\nabla_z= \p/\p z$. The coefficient   $A_{\rm s}$  is defined by  
\be 
A_{\rm s}= \ve_1^2/{\bar\ve} \cong  \ve_1^2/{\ve_0} .  
\label{eq:3.14}
\en 
which is small 
for pure fluids ($= (\ve_0-1)^2(\ve_0+2)^2/9\ve_0$ 
from  Clausius-Mossotti),  
  but can be  $10-100$ 
for polar binary mixtures.  
The  $F_{\rm dip}$ 
 is positive-definite 
for the fluctuations 
varying along ${\bi E}_0$ 
but vanishes for those varying 
perpendicularly to   ${\bi E}_0$. 
Thus  
$F_{\rm dip}$  produces  no shift of $T_{\rm c}$ 
in the mean-field theory, but 
its suppresses  the   fluctuations 
leading  to a fluctuation contribution 
to the shift  as  in (3.23) below.

Far from the capacitor plates we may set  
 $G({\bi r}, {\bi r}') = 1/4\pi |{\bi r}- {\bi r}'|$. 
Picking up 
only  the Fourier components  $\psi_{\small{\bi q}} 
= \int d{\bi r} \psi({\bi r}) 
{\rm e}^{i\small{\bi q}\cdot\small{\bi r}}$ with 
$qL \gg 1$, we obtain  
\be 
{F}_{\rm dip} 
={\frac{1}{2}}k_{\rm B}T_{\rm c} 
 g_{\rm e} \int_{\bi q} \hat{q}_z^2   
|\psi_{\small{\bi q}}|^2, 
\label{eq:3.15}
 \en
where $\hat{q}_z= q_z/q$. 
The strength of the interaction is given by 
\be
g_{\rm e} = (4\pi k_{\rm B}T_{\rm c})^{-1} A_{\rm s} {E_0}^2.
\label{eq:3.16}
\en 
A similar dipolar interaction    
 is well-known for  uniaxial ferromagnets  
\cite{dipolar1,Ah}.  
At the critical density (or composition) in one-phase states, 
the  structure factor $I({\bi q})$ in the presence of 
${F}_{\rm dip}$ becomes uniaxial as  
\be 
 I({\bi q})=  [ a_0(\tau-\tau_{\rm c}) + 
g_{\rm e} \hat{q}_z^2 + K_0 q^2]^{-1},   
\label{eq:3.17}
\en   
where $\tau_{\rm c}$ is a shift induced by electric field. 
For $g_{\rm e} \hat{q}_z^2>a_0\tau_{\rm c}$, the 
intensity decreases with increasing 
electric field even if $\tau_{\rm c}>0$.

(i) {\it Weak field regime}: 
If  $g_{{\rm e}}$ is  smaller than $a_0 \tau={\hat \chi}^{-1}$, 
the electric field is weak 
and  the  structure factor may be expanded  as 
\be 
I({\bi q})= I_0(q)  + I_0(q)^2 
 (a_0\tau_{\rm c}-g_{\rm e} \hat{q}_z^2) +\cdots,  
\label{eq:3.18}
\en 
where $ I_0(q)= 
{\hat \chi} /(1+q^2\xi^2)$ is the intensity 
at zero electric field with ${\hat \chi}  =I(0)$. 
 The {\it apparent}  shift 
arises from the bilinear term 
($\propto \psi^2$) 
in  $F_{e0}$ in (3.12):   
\be 
a_0 \tau_{\rm c}  =  
{\ve_2}{E}_0^2/4\pi k_{\rm B}T_{\rm c}   
= {\ve_2 }{g_{\rm e}}/A_{\rm s}.  
\label{eq:3.19}
\en 
Note that $a_0 \tau_{\rm c}$ and 
hence (3.18) are independent of $\Lambda$. 
The usual   critical behavior at long wavelengths 
$q\xi<1$ follows for    
 $\Lambda\xi \sim 1$, where   
 $\xi=\xi_{+0}\tau^{-\nu}$ is  the  correlation  
length. The asymptotic   scaling relations are   
${\hat \chi}= \Gamma \tau^{-\gamma}$ 
(or $a_0= \Gamma^{-1} \tau^{\gamma-1}$) and  
$K_0= {\hat \chi}^{-1}\xi^2= 
\Gamma^{-1}\xi_{+0}^2 \tau^{-\nu\eta}$,  where $\Gamma$ 
is a constant, and   
$\gamma\cong 1.24$, $\nu\cong 0.625$, 
and $\eta=0.03-0.05$ are the usual 
critical exponents.

(ii) {\it Strong field regime}: 
The electric field is strong for 
 $\tau-\tau_{\rm c}<\tau_{{\rm e}}$.  
The crossover  reduced 
temperature $\tau_{{\rm e}}$ and  the   characteristic 
wave number $k_{{\rm e}}$ are written as  
\be 
\tau_{{\rm e}}= (\Gamma g_{{\rm e}})^{1/\gamma}, \quad 
k_{{\rm e}}= \xi_{+0}^{-1} (\Gamma g_{{\rm e}})^{\nu/\gamma}. 
\label{eq:3.20}
\en 
As a rough estimate, 
we set $\Gamma \sim \xi_{+0}^3$, $\xi_{+0}\sim  
2.5\times 10^{-8}$cm, and 
$T_{\rm c}\sim  300$K to obtain 
$\tau_{\rm e} \sim 10^{-6} A_{\rm s}^{0.8} E_0^{1.6}$ 
with $E_0$ in units of $10^4$V$/$cm.  There is no experimental 
report in this regime. 
According to 
  the renormalization group theory 
in the presence of the uniaxial dipolar 
interaction (3.15) (which treats $F_0+F_{\rm dip})$\cite{Ah}, 
 there is no renormalization of the coefficient $g_{\rm e}$ 
and the dipolar interaction
  strongly suppresses the fluctuations  
with wave numbers smaller than $k_{\rm e}$  
 in the temperature range $\tau-\tau_{\rm c}<\tau_{{\rm e}}$  
resulting in   mean field critical behavior 
for  $d \ge d_{\rm c}=3$.  For these fluctuations we may  set 
  $\Lambda \sim k_{{\rm e}}$ to obtain  the renormalized coefficients, 
\be 
a_0 = \Gamma^{-1} \tau_{{\rm e}}^{\gamma-1}, \quad 
K_0= \Gamma^{-1}\xi_{+0}^2 \tau_{{\rm e}}^{-\nu\eta}, \quad 
u_0= g^* K_0^2k_{{\rm e}}^\epsilon /K_d ,
\label{eq:3.21}
\en     
where  
$K_d=(2\pi )^{-d}2\pi^{d/2}/\Gamma (d/2)$,  
   $\epsilon=4-d$, 
and $g^*=\epsilon/9+O(\epsilon^2)$ is the fixed point value 
of the  coupling constant 
$g=K_d u_0/K_0^2\Lambda^\epsilon$ 
 \cite{Onukibook}. The  shift 
$\tau_{\rm c}$ in  (3.17) in  strong  field   
consists of two terms  as 
$\tau_{\rm c}= \tau_{{\rm c}1}+\tau_{{\rm c}2}$. 
The origin of $\tau_{{\rm c}1}$  is  the same as that of 
 $\tau_{{\rm c}}$  in (3.19). Here using  (3.20) and (3.21) we obtain  
\be 
\tau_{{\rm c}1}=  {\ve_2}E_0^2/4\pi a_0k_{\rm B}T 
= (\ve_2/A_{\rm s}) \tau_{{\rm e}}.  
\label{eq:3.22}
\en  
The $\tau_{{\rm c}2}$ 
arises    from the quartic term ($\propto 
u_0$) in $F_0$ in (3.2):
\be
 \tau_{{\rm c}2}
= -\frac{3}{a_0} u_0  \int_{\bi q} \bigg [ 
\frac{1}{g_{{\rm e}}\hat{q}_z^2+ K_0 q^2}  - \frac{1}{K_0q^2} \bigg ] 
= 3g^*C_d\tau_{{\rm e}},  
\label{eq:3.23}
\en 
where   $C_d = K_d^{-1}  \int_{\bi \ell} 
\ell_z^2/\ell^2(\ell^4+\ell_z^2)$ 
($= \pi/4$ at $d=3$).  The   $\tau_{{\rm c}2}$ 
  is a positive 
fluctuation contribution meaningful 
 only in strong field. (A similar 
 shift was  calculated for 
near-critical fluids in  shear flow \cite{Onukibook}.)   
Therefore both $\tau_{{\rm c}1}$ 
and $\tau_{{\rm c}2}$ are of order $\tau_{{\rm e}}$. 
We also consider  the 
 correlation function 
$g({\bi r}) = \int_{\bi q} I({\bi q})
\rm{e}^{i{\small{\bi q}}\cdot{\small{\bi r}}}$ 
in  strong field. From    (3.17) and (3.21)  
 the scaling form  
$
g({\bi r})= G(x/\xi_\perp, y/\xi_\perp, z/\xi_\parallel)
/K_0 \xi_{\parallel}
$
holds with  
\be 
\xi_\perp = k_{\rm e}^{-1}  [\tau_{\rm e}/(\tau- \tau_{\rm c})]^{1/2}, 
\quad \xi_\parallel= k_{\rm e}\xi_\perp^2= 
k_{\rm e}^{-1}  \tau_{\rm e}/(\tau- \tau_{\rm c})>\xi_\perp.
\label{eq:3.24}
\en 
Here $G(0,0,Z) \sim Z^{-3}$ for $Z\gg 1$. The 
large-scale critical fluctuations 
($ \gg k_{\rm e}^{-1}$) are elongated 
along the $z$ axis, 
as expected in experimental papers $[13]$.

We now comment on previous work.\\
(i) If   $a_0$ is treated  as a constant, 
$T_{\rm c}\tau_{{\rm c}1}$  is the mean field shift 
in   Landau-Lifshitz's book  
for pure fluids \cite{Landau4}. The same shift was later   
proposed  for binary mixtures \cite{Beaglehole}.\\ 
(ii) 
 Debye and Kleboth \cite{Debye-Kleboth} 
 derived a reverse shift 
($\Delta T_{\rm c}=  - T_{\rm c}\tau_{{\rm c}1}$) 
 neglecting the inhomogeneity
 of $\bi E$ and setting 
${\bi E}={\bi E}_0$ in (3.7). The dipolar interaction was 
 nonexistent, leading to  
 the normalized turbidity change 
 ${\cal T}(E_0) /{\cal T}(0)-1\cong 
\hat{\chi}a_0\tau_{\rm c}$ for $k\xi \ll 1$, 
where ${\cal T}(E_0)$ is the turbidity in electric field 
 (see (1.7)).   
They found  turbidity  decreases  
in   nitrobenzene+ 2,2,4-trimethylpentane 
to  obtain   $\tau_{\rm c} 
= - 0.5\times 10^{-4}$ at $E_0= 45$kV$/$cm 
in agreement with their theory, 
 where   Hildebrand's 
 theory  gave $a_0= (4.3{\rm \AA})^{-3}$ and  
 data of $\ve(\phi)$ 
 as a function of the volume 
fraction $\phi$ of nitrobenzene  gave 
$\p^2 \ve/\p \phi^2= 28.7$. 
(Notice that their $|\tau_{\rm c}|$  is  of order 
$\tau_{\rm e}$ in (3.20)  if  $A_{\rm s} \sim 10$.) 
Subsequent light scattering  experiments 
 detected similar suppression  
in  a  near-critical binary mixture \cite{Ikushima} 
and in a polymer solution 
\cite{Fuller_e1}.    In our theory    (3.18) and (3.19) hold 
in weak field, so  for  $k\xi \ll 1$ we have  
\be 
{\cal T}(E_0) /{\cal T}(0)-1 \cong  
\hat{\chi}
 (a_0\tau_{\rm c}- g_{\rm e}/5) 
\quad {\rm or}\quad   \hat{\chi}
 (a_0\tau_{\rm c}- 2g_{\rm e}/5)
\label{eq:3.25}
\en  
for the polarization along the $z$ or $y$ axis. 
Thus we predict that the turbidity decreases  for 
$g_{\rm e}/5>a_0\tau_{\rm c}$ or for 
$A_{\rm s}>5\ve_2$ for any polarization. 
Here we expect that $A_{\rm s}$ is considerably 
larger than $|\ve_2|$ in polar mixtures 
(see the end of Subsection 3.3 also).    
Complex effects   also arise  from 
   a small amount of ions  which are present 
 in  most binary mixtures (see  Section 4).

\subsection{Macroscopic dielectric constant}@

We calculate 
the macroscopic dielectric constant 
$\ve_{\rm eff}= 
4\pi \av{Q}/S E_0 
= \av{D_z}/E_0$ in the absence of charges.  From 
the charge-free condition $\nabla \cdot \av{\bi D}= 
\nabla_z \av{D_z}=0$, 
 $\av{D_z}$ is homogeneous in the fluid as   
\be 
\av{D_z}=  \av{\ve}E_0 - \av{\delta\ve\nabla_z\delta\phi}
 = {\rm const.}
\label{eq:3.26}
\en  
Far from the boundary surfaces,  by setting $M=\av{\psi}$, 
we   obtain  
\be 
\ve_{\rm eff}= \ve_{0} + \ve_1 M + \frac{1}{2}\ve_2 
M^2  + \int_{\bi q} \bigg [ 
\frac{1}{2}\ve_2 -  A_{\rm s} 
 {\hat{q}_z^2} \bigg ]  I({\bi q}) ,  
\label{eq:3.27}
\en 
where the last term is   the $zz$ component of the tensor in (1.3).

(i) {\it Linear response}: 
The linear dielectric constant 
$\ve_{\rm eff}(0)$ is defined in the limit $E_0\rightarrow 0$, where 
 $\hat{q}_z^2$ in (3.27) may be replaced by   $1/3$. 
 Because $a_0 \psi^2$ corresponds to 
the energy density  in the 
corresponding Ising model (see the sentences below (1.2)), 
the renormalization 
group theory  gives 
 $\av{a_0 \psi^2}
 = {\rm const}.-   A \tau^{1-\alpha}/(1-\alpha) +\cdots 
$ with   $\alpha\cong 0.11$ 
(the specific-heat critical exponent)  and  
$A= R_\xi^d/(\alpha \xi_{+0}^{d})$  
 at the critical density or composition 
above $T_{\rm c}$ \cite{Onukibook}.  
In terms of this $A$,  the  specific heat  
$C_V$ for  pure fluids  and 
$C_{pc}$ for binary mixtures per unit volume grow 
as $- k_{\rm B}\p \av{a_0 \psi^2}/\p \tau 
\cong k_{\rm B}A\tau^{-\alpha}$.   
The  $R_\xi^d= \alpha A\xi_{+0}^d (\cong K_d/4)$ 
is a universal number (the two-scale-factor 
universality). It follows that 
$\ve_{\rm eff}(0) $ should    exhibit   weakly  singular behavior 
$[4$-$6,8]$, 
\be 
\ve_{\rm eff} (0) = \ve_{\rm c} +  D_1  \tau^{1-\alpha} + 
D_2 \tau + \cdots.     
\label{eq:3.28}
\en      
In our theory  $D_1= (A_{\rm s}- 3\ve_2/2)A/3a_0$ 
 depends  on the arbitrary cut-off
 $\Lambda$ since $a_0 \propto \Lambda^{(\gamma-1)/\nu}$ 
for $\Lambda \gg \xi^{-1}$, indicating inadequacy of our 
theory at short wavelengths. 
To calculate $D_1$  correctly 
we need to interpolate 
the renormalization group theory 
to a  microscopic theory \cite{Langer}. 
 On the other hand, 
the refractive 
index $n(\omega)=\ve(\omega)^{1/2}$ 
 at optical frequency  has also been predicted to 
be of  the  form of (3.28) \cite{Stell,Sengers_e}. 
However, despite a number of experiments, 
unambiguous detection of the weak singularity  
 in these quantities 
has been difficult for both pure 
fluids and binary mixtures \cite{Beysens}.

\begin{figure}[t]
\vspace{-1cm} 
\epsfxsize=3.80in 
\centerline{\epsfbox{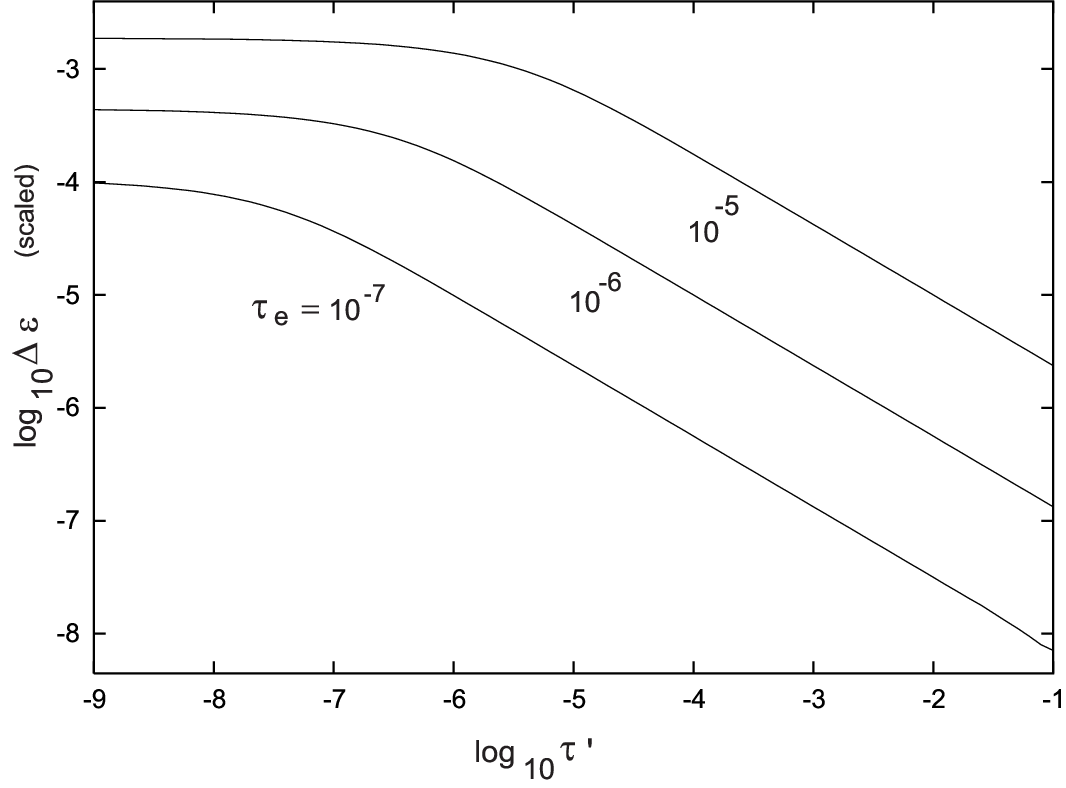}}
\caption{\protect%\narrowtext
\small{Nonlinear dielectric constant $\Delta \ve_{\rm eff}(\tau,E_0)$ 
(divided by a constant $B_0$) 
vs  reduced temperature $\tau'= 
\tau-\tau_{\rm c} >0$  in applied electric field 
$E_0$ for $\tau_{\rm e}= 10^{-5}, 10^{-6},$ and $10^{-7}$, 
where $\tau_{\rm e} \propto E_0^{1.6}$ from (3.20).  
  }} 
\label{2}
\end{figure}

(i) {\it Nonlinear response}: 
The nonlinear dielectric constant 
 $\Delta\ve_{\rm eff}= 
\ve_{\rm eff}(E_0)-   \ve_{\rm eff}(0)$ 
has been observed to become
 positive and grow near the critical point 
in  polar binary mixtures $[12$-$15]$.
Because small-scale 
fluctuations are insensitive to  electric field, 
such critical anomaly arises from 
nonlinear effects at  long wavelengths 
($q\ls \xi^{-1}$ for weak field and 
$q\ls k_{\rm e}$ for strong  field). 
Its calculation is therefore much  easier   
than that of $\ve_{\rm eff}(0)$. 
Use  of the structure factor (3.17) yields    
\bea
\Delta\ve_{\rm eff} &\cong& 
\frac{1}{16\pi}   (2A_{\rm s}/5- \ve_2+\ve_2^2/A_{\rm s}) 
K_0^{-2} g_{{\rm e}}   
\xi \quad  (\tau \gg \tau_{{\rm e}})
\label{eq:3.29} \\
&\cong &\frac{1}{16\pi}
(A_{\rm s}- \ve_2) K_0^{-1}k_{{\rm e}} 
 \hspace{2.3cm}   (0<\tau' \ls \tau_{{\rm e}}), 
\label{eq:3.30}
\ena 
In the first line we have   
 $K_0 \propto \xi^{\eta}$ and  
 $\Delta\ve_{\rm eff} \propto E_0^2 \xi^{1-2\eta}$, while 
in the second line 
 $K_0 \propto k_{\rm e}^{-\eta}$ 
 and   $\Delta\ve_{\rm eff}\propto E_0^{(1+\eta)/(2-\eta)}$.  
Here we set $\tau'= \tau-\tau_{{\rm c}}>0$.
Then  $\tau' \cong \tau$ for  weak field. 
If $\Delta\ve_{\rm eff}$  is  positive for any $\tau'$, 
we need to require $A_{\rm s}> \ve_2$. 
Previously   we calculated 
(3.29) without the terms   involving  
$\ve_2$ \cite{Onuki-Doi4}. 
For general $\tau'/\tau_{\rm e}$ we find  
the scaling,  
\be 
\Delta \ve_{\rm eff} 
=B_0  \tau_{\rm e}^{\nu(1+\eta)}  {F}_{\rm non}(k_{\rm e}\xi),  
\label{eq:3.31}
\en 
where $B_0$ is a constant, 
${F}_{\rm non}(x) \sim x^{1-2\eta}$ for $x \ll 1$,  
and   ${F}_{\rm non}(x) \rightarrow 1$ for $x \gg 1$. 
Here $x=k_{\rm e}\xi\cong  (\tau_{\rm e}/\tau')^\nu$.   
If we neglect the contributions involving  $\ve_2$ and 
set $\eta=0$, ${F}_{\rm non}(x)$ is calculated in a universal form,  
\be   
{F}_{\rm non}(x)= \bigg ( \frac{1}{x}+ 
\frac{1}{2x^{3}} \bigg ) \sqrt{1+x^2} 
-\frac{1}{2x^{4}}  \ln ({x}+\sqrt{1+x^2}) 
 -\frac{4}{3x}.
\label{eq:3.32}
\en 
In Fig.2 we plot $\Delta \ve_{\rm eff}/B_0= \tau_{\rm e}^\nu 
{F}_{\rm non}( k_{\rm e}\xi)$ vs $\tau'$ by neglecting $\eta$. 
For weak field  $\tau_{{\rm e}}\ll \tau$, 
we predict  $\Delta\ve_{\rm eff}\propto 
\tau^{-z_{\rm non}}E_0^2$ at the critical density or 
composition,  where $z_{\rm non}= \nu(1-2\eta)
= \gamma-2\beta \cong 0.58$ 
using  
$\gamma+2\beta=2-\alpha= 3\nu$  
\cite{Stell,Goulon,Onuki-Doi4}. 
 For  strong field 
$\tau_{{\rm e}}\gs \tau'$,     
$\Delta\ve_{\rm eff}$ should saturate into 
(3.30).

In  the experiments,  $z_{\rm non}$ 
was  around 0.4  and 
the coefficient $\Delta \ve_{\rm eff}/[\tau^{-z_{\rm non}}E_0^2]$ 
was proportional to $(\ve_{\rm A}-\ve_{\rm B})^4/\ve^2$ for 
various binary mixtures \cite{Rzoska,Rzoska1}.  
The latter aspect is in accord with 
(3.29) if $\ve_1 \sim \ve_{\rm A}-\ve_{\rm B}$ 
and $A_{\rm s}- 5(\ve_2-\ve_2^2/A_{\rm s})/2 
 \sim A_{\rm s}$. 
We also note that a weak tendency of saturation of 
$\Delta\ve_{\rm eff}$ was detected 
at   $\tau=3\times 10^{-4}$ 
with increasing  $E_0$ above $60$ kV$/$cm,   
 but  it was attributed to a negative 
critical temperature shift \cite{Orz}.   
These results  and Debye-Kleboth's data  suggest that 
$|\ve_2|$ is considerably smaller than  
 $A_{\rm s}$ at least in polar mixtures.

\subsection{Critical electrostriction}

We consider equilibrium of pure fluids in which  
$\av{{\bi E}} = {\bi E}_0({\bi r})$ 
varies  slowly in space. 
In equilibrium 
the chemical potential defined by 
the following is homogeneous \cite{Landau4}, 
\be  
\mu(\tau,M,E_0)= \av{\delta  F/\delta  n}
= \mu(\tau,M,0)  
- \frac{1}{8\pi n_{\rm c}}\ve_1 E_0^2= {\rm const}. 
\label{eq:3.33}
\en  
Here we set $\psi= (n-n_{\rm c})/n_{\rm c}$.   When ${\bi E}_0$ 
varies slowly compared with the correlation length,
  an  inhomogeneous average density variation 
is  induced as  
\be  
\av{\delta n}({\bi r}) \cong n_{\rm c} K_T 
{\ve}_1 E_0({\bi r})^2/8\pi  ,
\label{eq:3.34}
\en  
where $K_T= (\p n/\p \mu)_T/n^2   
\sim \tau^{-\gamma}/k_{\rm B}T_{\rm c} n_{\rm c}$  
is the isothermal compressibility. 
Experimentally,    the above relation    
 was   confirmed  optically for SF$_6$ 
 around a wire conductor 
\cite{Maryland},  
and  was used to determine 
$\mu(T,M)$ for $^3$He in a cell within 
which a parallel-plate capacitor was 
immersed \cite{Barmartz_el}.

This problem should be of great importance  on 
smaller scales particularly in near-critical polar binary 
mixtures.  For example, let us consider a 
spherical  particle 
 with charge $Ze$ and radius $R$
placed at the origin of the reference frame. 
The fluid is in a one-phase state 
 and is at the critical density or composition 
far from the particle, so $\tau>0$.  
In the Ginzburg-Landau 
scheme (3.2) and (3.9) yield 
\be 
(a_0\tau +u_0\psi^2 -K_0\nabla^2)\psi =h(r)\xi_{+0}^{-3}  
\cong 
A_0  \xi_{+0} r^{-4},   
\label{eq:3.35}
\en 
where $A_0= \ve_1 Z^2e^2 /8\pi\ve_0^2
k_{\rm B}T\xi_{+0}$.   The  $h(r)= 
A_0 (\xi_{+0}/r)^4$ is 
a  dimensionless space-dependent 
ordering field.  We may assume 
$\ve_1>0$ and $A_0>0$. 
 For   polar binary mixtures and/or for colloidal 
particles with $Z\gg 1$, 
$A_0$  can well exceed 1 (see Section 4). 
In the space region, where  
 $\xi< (\p\ln h(r)/\p r)^{-1} \sim r$, 
the usual scaling relations hold locally at each point 
and the renormalization yields 
the coefficients $a_0$, 
$u_0$, and $K_0$ dependent  on  $\tau$ and $h(r)$ 
\cite{Onukibook}.  
However,  $\xi<r$ is  
violated at small $r$ for $\xi>R$, where the 
gradient term in (3.35) 
($\propto \nabla^2\psi$)  is indispensable.   
The  linear relation $\psi \cong 
\Gamma \xi_0^{-3} \tau^{-\gamma} h(r)$ given in  (3.34) 
holds for $h(r) < \tau^{\beta\delta}$ 
and  $r>\xi$, where the former condition 
is rewritten as 
$r>\xi_{+0}A_0^{1/4}  \tau^{-\beta\delta/4}$.  
As $\tau \rightarrow 0$, 
the profile $\psi(r)$ becomes  very complicated 
depending on $A_0$, $\xi/R$, and the boundary condition at $r=R$.
Detailed discussions will appear shortly.

\subsection{Electric birefringence and dichroism} 
\begin{figure}[t]
\vspace{-1cm}
\epsfxsize=3.6in
\centerline{\epsfbox{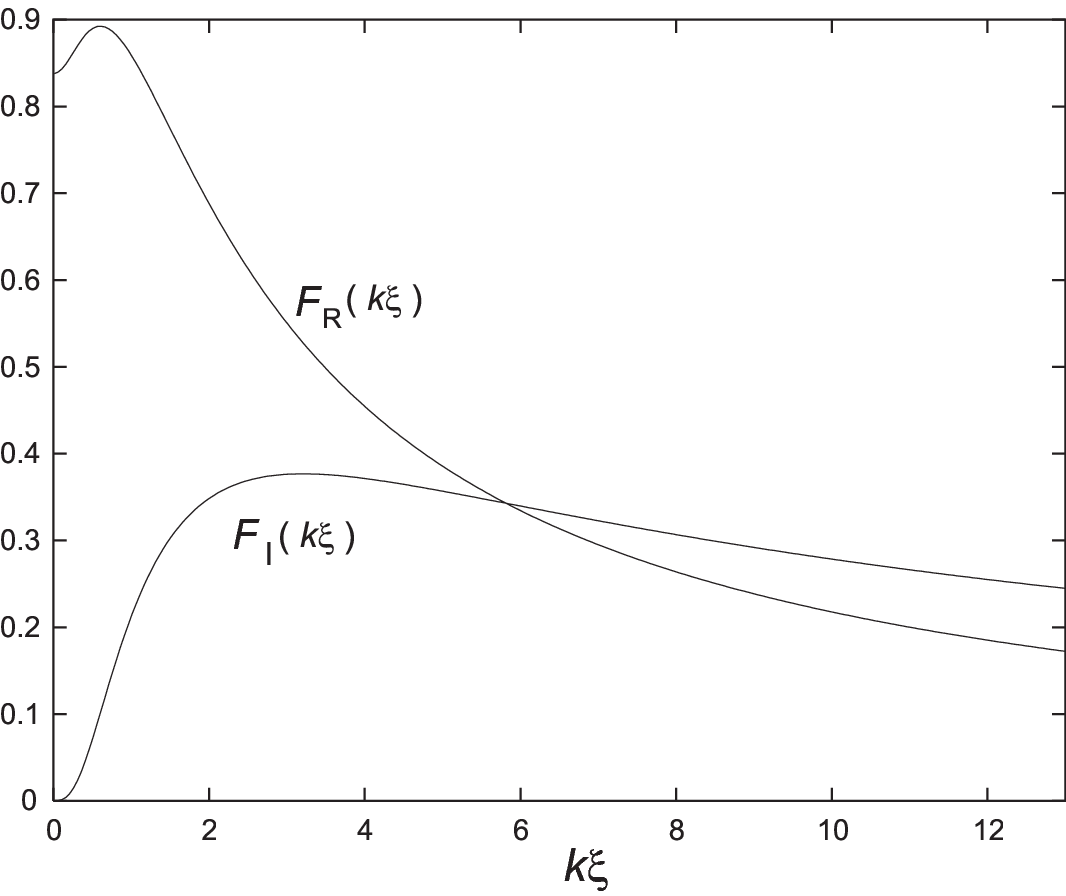}}
\caption{\protect%\narrowtext
\small{ Two scaling functions corresponding to form birefringence 
and dichroism \cite{Onuki-Doi4}. 
The $k$ is the wave number of probing light and 
$\xi$ is the correlation length. }}
\label{3}
\end{figure}
\begin{figure}[t]
%\hspace{0.8cm}
\vspace{-0.5cm}
\epsfxsize=6.3in 
\centerline{\epsfbox{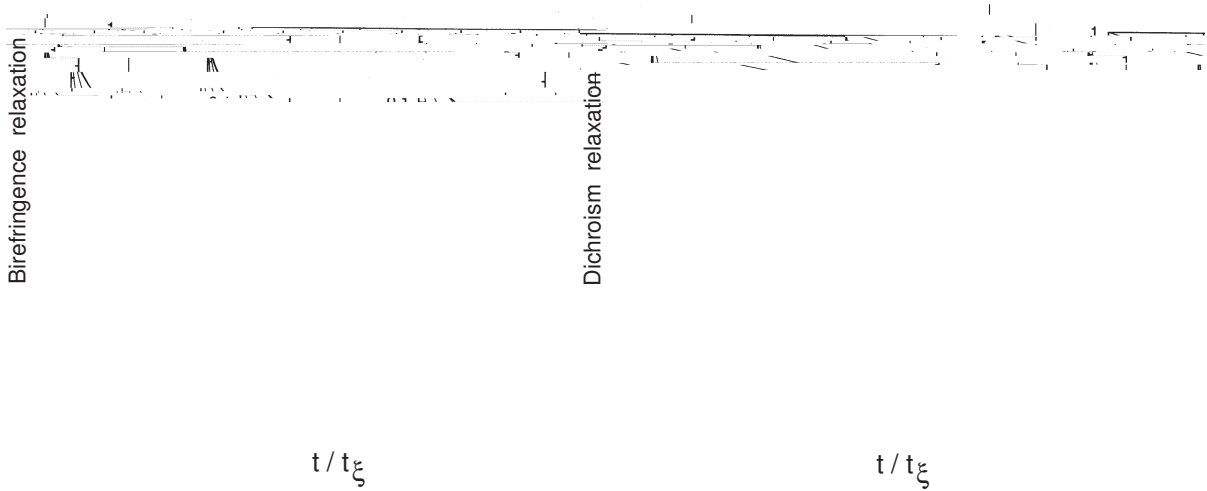}}
\caption{\protect%\narrowtext
\small{Normalized birefringence 
${\rm Re}\Delta n(t)/{\rm Re}\Delta n(0)$ 
and dichroism   
${\rm Im}\Delta n(t)/{\rm Im}\Delta n(0)$ 
vs $t/t_\xi$ on semilogarithmic scales 
for  $k\xi= 0.1,0.5,1$, and 2. 
} }
\label{4}
\end{figure}
 The   anisotropy   of the structure factor in (3.17) 
 has not yet been  measured in near-critical fluids, 
 but it gives rise to critical anomaly in 
 electric birefringence (Kerr effect) 
$[16$-$19]$
%\cite{Kerr0}-\cite{Kerr3}  
 and dichroism.   These effects 
 can be sensitively detected even in  the  weak 
field regime  $\tau _{\rm e} \ll \tau$  and even for not large 
$\ve_1$ using high-sensitivity optical techniques. 
We assume that  a laser beam with  optical 
frequency $\omega$ is passing through a near-critical fluid 
along the $x$ axis, while an  
electric field $E_0$ is applied along the $z$ axis.  
In  (1.5) we have $\Delta\ve_{yz}=0$ and 
 the difference $\Delta n= 
 (\Delta\ve_{zz}-\Delta\ve_{yy})/2n(\omega)$ 
 is written as \cite{Onuki_bi1,Onuki_bi2}
\be  
\Delta n= \frac{A_{{\rm op}}}{2n(\omega)} 
 \int_{\bi q} \frac{q_y^2-q_z^2}{q^2-k^2-i0} 
I({\bi k}-{\bi q}) ,
\label{eq:3.36}
\en 
where $k= \omega/n(\omega)$ is 
the laser wave number in the fluid and 
\be 
A_{{\rm op}}=  (\partial \ve(\omega)/\partial \psi)^2/\ve(\omega)  .  
\label{eq:3.37}
\en 
Here $A_{{\rm op}}$ 
is  different from the static coefficient $A_{\rm s}$ in (3.14). 
In fact $A_{\rm op} \ll A_{\rm s}$ for 
polar fluid mixtures where  
 $\ve (0) \gg \ve(\omega)$ at optical frequency  $\omega$ 
\cite{Is}. When $\tau \gg \tau_{{\rm e}}$, substitution of 
 (3.18) into  (3.36) gives   the steady state 
result,    
\be
\Delta n= \frac{A_{{\rm op}}}{32\pi^2n(\omega)} K_0^{-2}
 g_{{\rm e}}\xi  
 \bigg [ F_{\rm R}(k\xi)+ iF_{\rm I}(k\xi)\bigg ], 
\label{eq:3.38}
\en  
where the two scaling functions are plotted 
in Fig.3 and  are given by 
\bea
F_{\rm R}(x) &=& \frac{\pi}{4x^2}\bigg (1+\frac{1}{4x^2}\bigg )
 \bigg [ \bigg ( \frac{3}{2x}+2x\bigg ) \tan^{-1}(2x) -3 \bigg ],  
\nonumber\\
F_{\rm I}(x) &=&  \frac{\pi}{4x}\bigg [ 
\bigg (1+\frac{1}{x^2}+\frac{3}{16x^2} \bigg ) 
\ln (1+4x^2) - \frac{3}{4x^2}- \frac{5}{2}  \bigg ]. 
\label{eq:3.39}
\ena   
For $x \ll 1$ we have $F_{\rm R}(x) \cong 4\pi/15$ 
and $F_{\rm I}(x) \cong \pi x^3/3$. In 
the long wavelength limit ($k\xi \ll 1$),   
we find 
 $\Delta \ve_{\rm eff}/n(\omega)\Delta n= 
3(A_{\rm s} - 5\ve_2/6) /A_{{\rm op}}$ from (3.29) 
and (3.38),  
so  $\Delta \ve_{\rm eff}$ and $\lim_{k\xi\rightarrow 0}
\Delta n$ should  have 
the same critical behavior.   Experimentally, 
however, 
$\Delta n \propto \tau^{-z_{{\rm op}}}$ 
with $z_{{\rm op}}  \cong 0.65-0.85>z_{\rm non}$ was  obtained  
\cite{Rzoska1,Kerr2}.
\begin{figure}[t]
%\vspace{-8.5cm}
\vspace{-7.8cm}
\epsfxsize=4in 
\centerline{\epsfbox{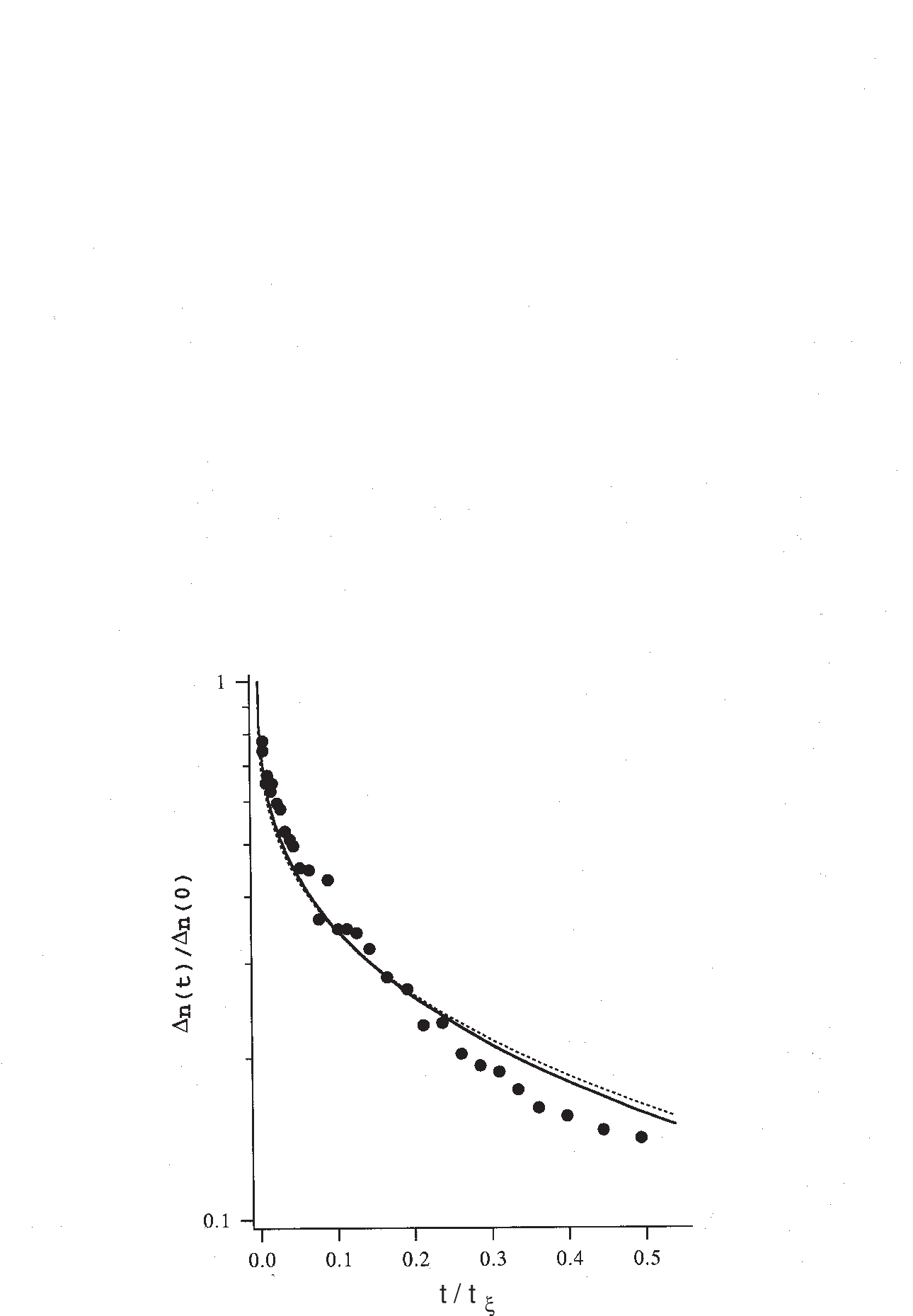}}
\caption{\protect%\narrowtext
\small {Comparison of the theoretical decay function 
(bold line) defined by (3.41) and data (dots) taken from Ref.\cite{Kerr2}. 
The stretched exponential function  $\exp(-2.3 x^{1/3})$ (dotted line) 
is a good approximation in the initial relaxation. 
} }
\label{5}
\end{figure}

 In transient electric birefringence,  
applied electric field is 
switched off (at $t=0$) and the subsequent relaxation 
 $\Delta n(t)$ (for $t>0$) is measured. 
For  near-critical binary mixtures,  
this experiment has  been carried out  
by applying a rectangular pulse of 
electric field $[16$-$19]$.  
If the pulse duration time $\Delta t$ 
is much longer than the relaxation time $t_\xi$ of the critical 
fluctuations, 
 the fluid can reach a   steady state while  
the field is applied.  
Here $t_\xi = 6\pi{\bar \eta}\xi^3/k_{\rm B}T_{\rm c}$  
with  ${\bar \eta}$ being  the shear viscosity.  
A remarkable finding is that 
$\Delta n(t)$ follows  a stretched exponential 
relaxation    at short times. 
We hereafter  explain our theory for weak field 
$\tau_{\rm e}\ll \tau $ \cite{Onuki-Doi4}.

We assume that the relaxation of 
 the structure factor  obeys  
\be 
I({\bi q},t)= I_0(q)  - I_0(q)^2 
 g_{\rm e} \hat{q}_z^2\exp(-2\Gamma_q t)  +\cdots ,
\label{eq:3.40}
\en 
where  $\Gamma_q= t_\xi^{-1} K_0(q\xi)$ is the relaxation rate 
with  $K_0(x)= (3/4)[ 1+x^2+(x^3- x^{-1}) \tan^{-1}x ]$ being  
the Kawasaki function \cite{Onukibook}.  Thus 
$\Gamma_q = Dq^2$ for $q\xi \ll 1$ with 
the diffusion constant 
$D=k_{\rm B}T_{\rm c}/ 6\pi{\bar \eta}\xi$.  
Here we neglect 
the term proportional to 
$\tau_{\rm c}$ in (3.18) because 
it does not contribute to $\Delta n$. 
We substitute (3.40) into (3.36) to obtain 
$\Delta n(t)$ as a function of $k\xi$ and $t/t_\xi$, where 
 the  initial  value $\Delta n(0)$ is 
 given by (3.38).  In Fig.4 we plot 
its normalized real and imaginary parts for various $k\xi$. 
The imaginary part decays slower than the 
real part. In the limit $k\xi\rightarrow 0$,  
it becomes real and 
is of the form, 
\be
\lim_{k\xi\rightarrow 0}
\frac{\Delta n(t)}{\Delta n(0)} 
=G(t/t_\xi)= \frac{4}{\pi} 
\int^\infty_0dy{y^2 \over (1+y^2)^2}\exp [-2K_0(y)t/t_\xi] ,  
\label{eq:3.41}
\en 
where the scaling function $G(x)$ behaves as 
\bea
G(x)&\cong& 1-2.3 x^{1/3} \cong \exp (-2.3x^{1/3}) 
  \quad\qquad  (x \ll  1)  \nonumber \\
&\cong& 0.2 x^{-3/2} ~\hspace{3cm}  \quad \quad\qquad   
(x \gg 1)  . 
\label{eq:3.42} 
\ena 
In Fig.5  data on 
butoxyetbaranol + water \cite{Kerr2}
 are compared with  (3.41)  for $t < t_\xi$. 
The  agreement  is excellent.  It 
demonstrates that  the theoretical $G(t)$ is 
nearly stretched-exponential  for $G(t)\gs 0.1$ 
or for $t/t_\xi \ls 0.5$.   
 As another theory, Piazza {\it et al.} 
 \cite{Kerr1} derived 
a stretched exponential decay of 
$\Delta n(t)$ on the basis of  a  
phenomenological picture on 
 the distribution of large 
clusters.

Let us consider the case $k\xi \ll 1$, where (3.41) 
is a good approximation for $t \ls  1/Dk^2$.   
At  $t \sim 1/Dk^2$,  
$G(t/t_\xi)$  becomes a very small 
number of order $0.2 (k\xi)^3$.   
In the later time region   
 $t \gs  (Dk^2)^{-1}$, (3.41) cannot be used and 
 the  fluctuations with wave numbers  
 of  order $(Dt)^{-1/2}$ 
give rise to the following birefringence signal, 
\be 
\frac{{\rm Re}\Delta n(t)}{{\rm Re}\Delta n(0)} 
 \cong  0.2 (t/t_\xi)^{-3/2} (Dk^2t)^{-1}  .
\label{eq:3.43}
\en 
Data of  birefringence relaxation 
 in Ref.\cite{Kerr3} indicated that 
the decay becomes faster than predicted by 
(3.41) at long times. The same tendency can be 
seen in Fig.4. 
In future   such data 
 should be  compared with 
the theoretical curves for finite $k\xi$. 
 We note that  transient electric 
dichroism  has not yet been measured  for 
near-critical fluids, but was 
 measured for a polymer solution\cite{Fuller_e2}.

\subsection{Interface instability induced by 
 electric field }

An interface 
between two immiscible fluids  becomes unstable 
against surface  undulations 
in   perpendicularly applied  electric field. 
This is because 
 the electrostatic energy is  higher  
 for  perpendicular  field  than   
for   parallel field. 
  In classic papers \cite{Frenkel}
 the instability on an  interface between 
conducting and nonconducting fluids (such as Hg and air) 
was treated.  In helium systems, where  
an  interface can be   charged with ions, 
the critical field is  much decreased  and  
intriguing surface patterns have been observed   \cite{Gorkov}.

We will  derive  this instability 
in the simplest manner \cite{Onuki_interface}. 
Let  a planar interface be placed at $z=z_0$ 
 in a near-critical fluid without ions in weak field,  
 $\tau<0$ and  
$|\tau| \gg \tau_{\rm e}$. 
 If the interface 
is displaced by a small height 
$\zeta (x, y)$, we may set 
${\p}\psi/\p z= \Delta\psi \delta(z-z_0 -\zeta)$ in (3.13), 
where $\Delta\psi$ is 
the order parameter  difference between the two phases. 
This means that  the effective surface charge density is given by 
$-\ve_1\Delta\psi E_0$. 
If $z_0$ is far from the boundaries, 
 we obtain  
\be
{F}_{\rm dip} =  \frac{1}{8\pi}   
C_{\rm e} {\int}d{\bi r}_{\perp}{\int}d{\bi r}'_{\perp}
  [|{\zeta ({\bi r}_{\perp})-\zeta 
({\bi r}'_{\perp})|^2
 + |{\bi r}_{\perp} -{\bi r}'_{\perp}|^2} ]^{-1/2},
\label{eq:3.44}
\en
where ${\bi r}_{\perp}=(x,y)$,   
${{\bi r}}'_{\perp}=(x',y')$,  and 
\be 
C_{\rm e}=  A_{\rm s} E_0^2  (\Delta\psi)^2/4\pi. 
\label{eq:3.45}
\en  
We expand the integrand of (3.44) 
in powers of 
$[\zeta ({\bi  r}_{\perp})-\zeta({{\bi r}_{\perp}}')]^2$. 
The first correction is negative and bilinear in $\zeta$. 
In terms of the Fourier transformation $\zeta_{\small{\bi k}}$ 
we have  
\bea
\Delta {F}_{\rm dip}  
&=&  - \frac{1}{16\pi}  C_{\rm e} 
{\int}d{{\bi r}}_{\perp}{\int}d{{\bi r}}'_{\perp}
\frac{ 
1}{ 
{|}{{\bi r}}_{\perp} - {{{\bi r}}_{\perp}}'{ |}^{3} } 
\left [\zeta ({\bi r}_{\perp})-
\zeta ({{\bi r}}'_{\perp}) \right ]^2 \nonumber\\
&=& -   \frac{1}{2} C_{\rm e}  
 {\int}_{\bi k}
k \mid\zeta_{\small{\bi k}}\mid^2, 
\label{eq:3.46}
\ena  
where  $\bi k$ is the two-dimensional wave vector with   
$k= |{\bi k}|$   and $\int_{\bi k}= 
(2\pi)^{-2} \int d{\bi k}$. We 
have used  the  formula 
${\int}d{{\bi r}}_{\perp}
[1-\exp(i{\small{\bi k}}\cdot{{\bi r}}_{\perp})]
/r_\perp^{3/2}=4\pi k$. Including gravity we write the 
total free energy change  
due to the surface deformation as  
\be 
\Delta F_{\rm surface}=\frac{1}{2} 
\int_{\bi k} [ \sigma k^2 +g \Delta\rho - C_{\rm e}k] 
\mid\zeta_{\small{\bi k}}
\mid^2,
\label{eq:3.47}
\en 
where $\sigma$ is the surface tension,  $g$ is the gravitational 
acceleration, and $\Delta \rho$ is the mass density difference 
between the two phases. The coefficient in front of 
$\mid\zeta_{\small{\bi k}}
\mid^2$ is minimum at $k= 
(C_{\rm  e}/2\sigma)^{1/2}$ and an   
instability is triggered for 
$C_{\rm  e} >2(\sigma g \Delta\rho)^{1/2}$. 
For polar mixtures this criterion 
is typically  $E_0> |\tau|^{\nu/2-3\beta/4}$kV$/$cm on earth.

\section{Near-critical fluids with ions} 
\setcounter{equation}{0}

We will discuss the effects of  ions doped in binary mixtures. 
It has long been 
 known that  even a small fraction  of ions (salt) 
 with $c \ll 1$  dramatically changes  the liquid-liquid 
phase behavior in  polar 
binary mixtures, where  $c$ is 
the mass or mole  fraction of ions. 
For small $c$, the UCST coexistence curve 
 shifts upward as 
\be 
(\Delta T)_{\rm c} = A_1 c + O(c^2)
\label{eq:4.1}\en 
with large 
positive coefficient $A_1$, expanding   
 the region of demixing. 
For example, $A_1/T_{\rm c} \sim 10$ 
with $T_{\rm c} \sim 300$K   when  
  NaCl  was added to  
cyclohexane$+$methyl alcohol \cite{polar1} 
and to triethylamine+water \cite{polar2}.  
Similar large impurity effects 
were observed when water was added to 
methanol+cyclohexane \cite{polar3}. 
In some polar mixtures, 
even  if they are  miscible at all $T$ 
at atmosphere pressure without salt, addition of  
a  small amount of  salt 
gives  rise to reentrant phase separation behavior 
 \cite{Kumar,Misawa}.

We consider  two species of ions  with charges, 
$Ze$ and $-e$, at very 
 low densities, $n_1({\bi r})$ and $n_2({\bi r})$, 
 in a near-critical 
binary mixture.  The average densities are 
written as $\av{n_1}_{\rm s}={\bar n}$ 
and $\av{n_2}_{\rm s}=Z{\bar n}$, where $\av{\cdots}_{\rm s}$ denotes 
taking the  space average. The charge neutrality 
condition yields $Z \av{n_1}_{\rm s}= \av{n_2}_{\rm s}$. 
The Debye wave number 
$\kappa$ and the Bjerrum length 
$\ell_{\rm B}$($=7\AA$ for water at $300$K) 
are defined  by  
\be 
\kappa= [4\pi(Z^2+Z) \bar{n} e^2/\ve k_{\rm B}T]^{1/2}, 
\quad 
\ell_{\rm B}= e^2/\ve k_{\rm B}T . 
\label{eq:4.2}
\en

\subsection{Ginzburg-Landau theory}

We set ${\bi p}= {\chi}(\psi) {\bi E}$ 
from the beginning.  Then 
the electric potential $\phi$ satisfies 
$-\nabla\cdot\ve\nabla{\phi}=4\pi\rho$  with $\ve=1+4\pi \chi$ 
and $\rho=e(Zn_1-n_2)$. 
We assume that  the free energy 
$F$ in (2.2) in the fixed charge condition is of the form, 
\be 
F = F_0\{\psi\}+ k_{\rm B}T \int d{\bi r}
\bigg [ n_1 \ln n_1 +   
n_2 \ln n_2    
+ (w_1n_1+ w_2n_2)\psi \bigg ] 
 + \frac{1}{8\pi}\int d{\bi r} \ve {\bi E}^2 , 
\label{eq:4.3}
\en 
where the first term depends only on 
$\psi$ and is given by (3.2), and the terms proportional to 
$w_1$ and $w_2$ arise from an  energy decrease   due 
to  microscopic polarization of the fluid around 
 individual ions.  In the neighborhood of  an ion of 
 species $\alpha$ $(\alpha=1,2$), 
 the electric field and the polarization 
are  given by  
${\bi E}_{\rm ind}=- \nabla (Z_\alpha e/\ve r)$ and 
${\bi p}_{\rm ind}= \chi {\bi E}_{\rm ind}$ in terms of 
the local $ \chi$ and $\ve$, where we write $Z_1=Z$ 
and $Z_2=-1$. The resultant electrostatic energy density 
is localized near the ion $(\propto r^{-4}$) and its 
space integral is  
$\int d{\bi r}\ve {\bi E}_{\rm ind}^2/8\pi= 
Z_\alpha^2e^2/2\ve r_{\alpha}$, 
where $r_\alpha$ is  the lower cutoff   
representing  an effective radius of an ion of 
 species $\alpha$. Here the screening length 
is assumed to be much longer than 
$r_\alpha$.    Due to the 
polarization,  the decrease of the 
 electrostatic energy 
(solvation free energy per ion) 
is given by the Born formula\cite{Is,Born}, 
\be 
\Delta E_\alpha=(1/\ve-1) Z_\alpha^2e^2 / 2 r_\alpha. 
\label{eq:4.4}
\en   
Neglecting $\psi$-dependence of $r_\alpha$ 
we estimate $w_\alpha$   as 
\be 
w_\alpha= (k_{\rm B}T)^{-1} 
\frac{\p}{\p \psi} (\Delta E_\alpha)=   
-\ve_1 Z_\alpha^2 \ell_{\rm B}/2\ve_0r_\alpha . 
\en   
The last term in (4.3) 
is the electrostatic free energy $\Delta F$
arising from slowly-varying fluctuations 
and depends  on the boundary condition.  
As a clear illustration, if  
 all the quantities are functions of $z$ only 
 in the charge-fixed condition, we have 
\be 
\Delta F= 2\pi S\int_0^L dz 
[\sigma_{\rm ex}-w(z)]^2/\ve(z), 
\label{eq:4.6} 
\en 
where $\sigma_{\rm ex}$ is the  capacitor charge 
density and  
$w(z)= \int_0^z dz' \rho(z')$. 
Though  neglected here,   
 electrostriction should also be 
investigated around charged 
particles (see Subsection 3.4), which can well 
produce a shift of the critical composition.

In this model the chemical potentials   
 $\mu_\alpha= \delta F/\delta n_\alpha$ of the ions 
are given by 
\be 
\mu_\alpha= k_{\rm B}T (\ln n_\alpha +1 + w_\alpha\psi)   
+Z_\alpha e \phi . 
\label{eq:4.7}
\en  
In equilibrium $\mu_\alpha$ become homogeneous, leading to   
ion distributions, 
\bea 
n_\alpha ({\bi r}) 
&=& n_{\alpha 0} \exp [ -w_\alpha \psi({\bi r}) 
 -Z_\alpha e\phi({\bi r})/k_{\rm B}T ] \nonumber\\
&\cong&
  n_{\alpha 0} [1-w_\alpha  \psi({\bi r}) 
 -Z_\alpha e  \phi({\bi r})/k_{\rm B}T ]. 
\label{eq:4.8}
\ena 
The second line holds when the exponents in the first line 
are  small.  Let us use  the second line to derive 
 Debye-H$\ddot{\rm u}$ckel-type relations.
Then  $n_{\alpha 0} \cong \av{n_\alpha}_{\rm s}
   [1+w_\alpha  \av{\psi}_{\rm s} 
 +Z_\alpha  e  \av{\phi}_{\rm s}
/k_{\rm B}T ]$.  The potential $\phi$ satisfies  
\be 
\ve_0 (\kappa^2-\nabla^2) \phi =\ve_1\nabla\psi\cdot\nabla\phi + 
{4\pi}e[ Zn_{10}-n_{20}-  Z{\bar n} (w_1-w_2)\psi]. 
\label{eq:4.9}
\en 
The right hand side consists of new contributions 
dependent on $\psi$, where the 
first  term  is important 
in the presence of applied field. 
  We here notice  the following.\\ 
(i) In applied electric field we can examine  
 the  ions distribution accumulated near the boundaries 
using  (2.4) and (4.8) (or (4.9)).\\
(ii) When $w_1\neq w_2$,   charge distributions 
arise around domains or wetting layers.  
Let the interface thickness ($\sim \xi$) be  
shorter than the Debye length 
$\kappa^{-1}$. For 
 a spherical domain with radius $R$,  
for example,  the   electric 
potential $\phi(r)$ is a function of the 
distance $r$ from the center of the sphere and then   
\be 
\Delta \phi=\phi (0)-\phi (\infty)= 
  [1-e^{-\kappa R}(1+\kappa R)]
\Phi_{\rm sat}. 
\label{eq:4.10}
\en  
For $\kappa R\gg 1$,  $\Delta \phi$ saturates 
into  
\be 
\Phi_{\rm sat}=- 
[k_{\rm B}T/e(Z+1)]  (w_1-w_2)\Delta\psi,  
\label{eq:4.11}
\en 
where $\Delta \psi$ is 
the difference of $\psi$ between the two phases. 
For $\kappa R\ll 1$  we find   
$\Delta \phi \cong (\kappa R)^2\Phi_{\rm sat}$. 
The  same potential difference $\Phi_{\rm sat}$  arises 
across a planar interface. 
For example, if $w_1-w_2 \sim 3$, $Z=1$,   and  
$\Delta\psi \sim 10^{-1}$, 
we have $|\Phi_{\rm sat}| \sim 10^{-2} $V.\\ 
(iii) In slow relaxation of $\delta \psi$, the deviations 
$\delta \mu_\alpha=
k_{\rm B}T ( \delta n_\alpha/\av{n_\alpha}_{\rm s} + w_\alpha \delta\psi)  
+Z_\alpha e \delta\phi$ should quickly relax to zero. 
Then the deviations 
$\delta n_\alpha$ are written in terms of $\delta\psi$ as    
\be 
\delta n_{1}+ \delta n_{2}= 
-{\bar n}( w_1+Z w_2) \delta\psi,
\label{eq:4.12}
\en 
\be  
(\kappa^2-\nabla^2) 
\rho =  \bigg [Ze\bar{n}  (w_1-w_2)\nabla^2
+\frac{1}{4\pi}\ve_1 \kappa^2  {\bf E}_0\cdot\nabla \bigg ]
\delta \psi, 
\label{eq:4.13}
\en  
where  we allow the presence of 
electric field $E_0$ to derive (4.22) below.

\subsection{Ion effects on  phase transition}

We consider   small  fluctuations 
in  a  one-phase state without electric field 
($E_0=0)$. The  fluctuation contributions  to $F$ 
in the  bilinear order 
are  written as 
\bea 
\frac{\delta F}{k_{\rm B}T}  &=& 
\int_{\bi q} \bigg [
\frac{1}{2}({a_0\tau+C_0 q^2})  |\psi_{\small{\bi q}}|^2 
+ \sum_{\alpha=1,2} \bigg (
\frac{|n_{\alpha{\small{\bi q}}}|^2}{2\av{n_\alpha}_{\rm s}}  + 
w_\alpha n_{\alpha {\small{\bi q}}}\psi_{\small{\bi q}}^*\bigg )   
+  \frac{2\pi }{k_{\rm B}T\ve_0 q^2}|\rho_{\small{\bi q}}|^2\bigg ]  
\nonumber\\
&=& \frac{1}{2}  \int_{{\bi q}} \bigg [
a_0\tau+C_0 q^2-  \frac{\bar{n}(w_1+Zw_2)^2 }{1+Z}
- \frac{Z\bar{n}(w_1-w_2)^2}{(1+Z)(q^2+\kappa^2)}{q^2}   \bigg ] 
  |\psi_{\small{\bi q}}|^2. 
\label{eq:4.14}
\ena 
In the first line 
 $n_{\alpha{\small{\bi q}}}$ and  
$\rho_{\small{\bi q}}$ are  the Fourier 
transformations of $\delta n_\alpha=n_\alpha-\av{n_\alpha}_{\rm s}$ 
and the charge density $\rho= e( Zn_1- n_2)$. 
In the second line we have expressed 
$n_{\alpha{\small{\bi q}}}$ 
in terms of $\psi_{\small{\bi q}}$ using  (4.12) and (4.13) and   
minimized   the first line.   
 We introduce a  parameter,  
\be 
\gamma_{\rm p}=  |w_1-w_2|/[(1+Z) (4\pi C_0\ell_{\rm B})^{1/2}]. 
\label{eq:4.15}
\en 
This number is independent of the ion density 
and represents the strength  of 
asymmetry in  the ion-induced polarization between   the 
two components. The 
structure factor at $E_0=0$ in the mean field theory 
is written as  
\be 
1/I_0(q)= a_0(\tau-\tau_{\rm ion}) 
+ C_0 q^2 [1- \gamma_{\rm p}^2\kappa^2/(q^2+\kappa^2)] , 
\label{eq:4.16}
\en 
where 
\be 
\tau_{\rm ion}
= (w_1+Zw_2)^2\bar{n}/(1+Z){a_0} .  
\label{eq:4.17}
\en

We draw the following conclusions.\\
(i) If $\gamma_{\rm p}<1$, 
$I_0(q)$ is maximum at $q=0$ and 
the  critical temperature shift due to ions 
is given by $T_{\rm c}\tau_{\rm ion}$ in the form of  
(4.1).  As a rough estimate for $Z=1$,  we set 
$a_0^{-1} \bar{n} \sim 
\xi_{+0}^3 \bar{n}  \sim {c}$, where 
$ c$ is the mass or mole  fraction. 
Then  $\tau_{\rm ion}
\sim  (w_1+w_2)^2 {c}$. If $|w_1+w_2| \sim 3$, this result 
is consistent with the experiments 
\cite{polar1,polar2}. In future 
experiments let $1/C_0I_0(q)$ vs $q^2$ be
 plotted; then,  
the slope is $1-\gamma_{\rm p}^2$ 
for $q \ll \kappa$ 
and is $1$ for $q \gg \kappa$. 
This changeover is  detectable unless $\gamma_{\rm p}\ll 1$.\\
(ii) The case $\gamma_{\rm p}=1$ 
corresponds to a so-called Lifshitz point \cite{Lu}, 
where $1/I(q)-1/I(0) \propto q^4/(q^2+\kappa^2)$.\\ 
(iii) If $\gamma_{\rm p}>1$, the structure factor 
attains a maximum at an 
intermediate wave number $q_{\rm m}$ given by  
\be 
q_{\rm m}= ( \gamma_{\rm p}-1)^{1/2} 
 \kappa \propto {\bar n}^{1/2}.
\label{eq:4.18}
\en 
The maximum structure factor  
 $I_0(q_{\rm m})$ diverges as  $\tau \rightarrow 
\tau_{\rm ion}'$,  where 
\be 
\tau_{\rm ion}'= \tau_{\rm ion} +  
(\gamma_{\rm p}-1)^2\xi_{+0}^2 \kappa^2 ,
\label{eq:4.19}
\en 
with the aid of  $C_0= a_0\xi_{+0}^2$.  
A  charge-density-wave 
phase should be realized for $\tau<\tau_{\rm ion}'$. 
It is remarkable that this mesoscopic phase appears however small $\bar n$ is 
(as long as $\gamma_{\rm p}>1$ and $q_{\rm m} L \gg 1$, 
$L$ being the system length).  
Here relevant is the coupling of the order parameter 
and the charge density in the form 
$\propto \psi \rho$ in the free energy density, which generally 
exists in ionic systems. 
This    possibility of 
a mesoscopic phase  was already  predicted  
for  electrolytes  \cite{metal}, 
but has not yet been confirmed in 
experiments.

Furthermore, we note that $\gamma_p$ 
increases with increasing $Z$. 
As an extreme case, 
we  may add  charged colloid particles 
with  $Z \gg 1$ and radius $R$ 
 in a near-critical polar mixture, 
where   $w_1$ grows as $Z^2$  from (4.5) and 
\be 
\gamma_{\rm p} \cong 
(\ell_{\rm B}/16\pi C_0)^{1/2}(\ve_1 /\ve_0)Z/ R. 
\label{eq:4.20}
\en 
Note that $Z\xi_{+0}/R$  
 can be made considerably 
larger than 1  with increasing $R$.  
Notice that the ionizable points on the surface 
is proportional to the surface area $4\pi R^2$.

\subsection{Nonlinear effects 
under  electric field}

 In most of the previous experiments, 
 a  pulse of strong electric field has been applied. 
For example, the field strength was $E_0= 10^4$V$/$cm 
and the pulse duration time was 
$\Delta t=1$ms at $T \sim 300$K 
\cite{Rzoska,Orz}. Let us set $Z=1$. 
As can be seen from the general relation 
(2.18),  we can neglect ion accumulation 
at the capacitor  plates  if 
\be 
\sigma_{\rm ex}= \ve E_0 /4\pi \gg 
e{\bar n}v_{\rm dri} \Delta t \quad {\rm or}\quad 
\Delta t \ll t_{\rm dri}.
\label{eq:4.21}
\en  
Here $v_{\rm dri}= \zeta^{-1}eE_0$  is the drift velocity,  
 $\zeta$ is the friction coefficient 
related to the  diffusion 
constant by $D_{\rm i}=k_{\rm B}T/\zeta$, and $t_{\rm dri}=
(\kappa^2 D_{\rm i})^{-1}$ 
is the drift time. For $\Delta t>t_{\rm dri}$ 
the electric field far from the 
boundaries are shielded if $\sigma_{\rm ex} <e{\bar n}L$. 
Typical experimental values 
are $D_{\rm i} \sim 10^{-6}$cm$^2/$s, 
$\zeta \sim 10^{-8}$g$/$s,  $\ve \sim 50$,  
and   $E_0= 10^4$V$/$cm, leading to   
  $v_{\rm dri} \sim 1$ cm$/$s. Then 
the condition (4.21) 
becomes ${\bar n}\Delta t  \ll 10^{12}$cm$^{-3}$s  and 
 can well hold in experiments \cite{Or_this}. 
Under this condition and in a one-phase state,   
the bulk region remains  homogeneous and 
the electric field is not yet shielded.  
If we are interested in slow motions 
of $\psi$, we may assume (4.12) and (4.13) to obtain 
the decay rate of $\psi_{\small{\bi q}}$ in the form
$\Gamma_{\small{\bi q}} =  
L_0 q^2/I({\bi q})$ 
($L_0$ being the kinetic coefficient)  with 
\be 
{I({\bi q}) }^{-1} = {I_0(q) }^{-1} 
+[  - {\ve_2} 
+  A_{\rm s} {q_z^2}/({q^2+\kappa^2})] (E_0^2/4\pi k_{\rm B}T)  , 
\label{eq:4.22}
\en 
where $I_0(q)$ is given by (4.16).   
If the pulse duration time $\Delta t$ is much longer 
than the relaxation time $1/\Gamma_{\small{\bi q}}$, 
the structure factor is given by 
$I({\bi q})$. Here   
${\hat q}_z^2= q_z^2/q^2$ in (3.17) is replaced by 
${q_z^2}/({q^2+\kappa^2})$ in the presence of ions due 
to the Debye screening. Let us consider 
the nonlinear dielectric constant in 
(3.29) and the electric birefringence  
$\lim_{k\xi\rightarrow 0}\Delta n$
in (3.36) on the order of $E_0^2$. 
For  $\kappa \xi <1$ the ion effect is small, but 
for  $\kappa \xi >1$ they  should 
 behave as 
\be 
\Delta\ve_{\rm eff}/A_{\rm s} \sim   
 \Delta n/A_{\rm op}   \propto E_0^2    
\kappa^{-1}.   
\label{eq:4.23}
\en 
This crossover occurs  for $c \sim {\bar n}\xi_{+0}^3 
 > \tau^{2\nu}$  in  the weak 
field regime  $\tau \gg \tau_{\rm e}$.

We also comment on the Joule heating. 
While the ions are drifting, 
the temperature increasing rate is given by 
\be 
C \frac{d}{dt}T = 2{\bar n}v_{\rm dri} eE_0= 
2{\bar n} e^2E_0^2/\zeta= \ve E_0^2/4\pi t_{\rm dri} ,  
\label{eq:4.24}
\en 
where $C$ is the specific heat per unit volume at constant volume 
and composition. 
For  $\Delta t>t_{\rm dri}$ the temperature will increase by 
$(\Delta T)_{\rm Joule} \sim \ve E_0^2/4\pi C$. 
By setting $C\sim k_{\rm B}\xi_{+0}^{-3}$, 
we obtain $d\tau/dt  
\sim  10^{-19}{\bar n} \sim 10^{-6}t_{\rm dri}^{-1}$  
s$^{-1}$  at  $E_0=
10^4$V$/$cm. The heating is negligible  
for very small ${\bar n}\Delta t$ or for 
not very small $\tau$.

As other nonlinear problems involving ions, we mention transient 
relaxation of the charge distribution 
after application of dc field, 
response to oscillating field, 
and effects of charges 
on wetting layers and 
interfaces between the two phases.

\section{Liquid crystals in electric field} 
\setcounter{equation}{0}

In liquid crystals  
near the isotropic-nematic transition,  
the order parameter is the symmetric, traceless, 
orientation tensor 
$\psi= {\{}Q_{ij}\}$ (which should be distinguished from 
the electric charge $Q$ on a capacitor plate). 
The polarization and the electric induction are 
written as ${\bi p}= \tensor{{ \chi}}\cdot {\bi E}$ 
 and  ${\bi D}= 
\tensor{\ve}\cdot{\bi E}$, respectively, 
 where  the polarizability tensor 
$\tensor{{\chi}}$   depends on $Q_{ij}$ and 
the local dielectric tensor $\tensor{\ve}=
1+4\pi \tensor{\chi}  
$ reads    
\be 
\ve_{ij}= \ve_0 \delta_{ij} + \ve_1 Q_{ij} +  
\ve_{21} Q^2  \delta_{ij} + \ve_{22} (Q^2)_{ij} +\cdots , 
\label{eq:5.1}
\en
where  $Q^2= \sum_{ij}Q^2_{ij}$ hereafter.  
In the nematic phase we may set $Q_{ij}=S(n_in_j -\delta_{ij}/3)$ 
in terms of the amplitude $S$ and the director   $\bi n$. 
Then, 
\be 
\ve_{ij}= \ve_0'(S) \delta_{ij}+ \ve_1'(S)  n_in_j, 
\label{eq:5.2}
\en 
where $\ve_0'= \ve_0 + S^2(2\ve_{21}/3+ \ve_{22}/9)$ 
and $\ve_1'= \ve_1S+ \ve_{22}S^2/3$. 
These relations are analogous to (1.2) for near-critical fluids. 
An  important difference is that the tensor 
${\{}Q_{ij}\}$ is not conserved and its  average is  sensitive 
to applied electric field, while the average order parameter in 
near-critical fluids is fixed or conserved.

\subsection{Pretransitional growth} 

Here we examine the effect of the field-induced dipolar 
interaction. 
If we assume 
no mobile  charges  inside the fluid, 
the free energy functional 
is given by  $F=F_0+ 
\int d{\bi r}{\bi E}\cdot{\bi D}/8\pi$ 
 at fixed capacitor charge, 
where  
\be 
F_0=  \int d{\bi r}[\frac{A(T)}{2}  Q^2+ 
\lambda \sum_{j}Q_{jj}+ \frac{L_1}{2}  \sum_{ijk} 
[\nabla_k Q_{ij}]^2+ \frac{L_2}{2}  \sum_{ijk} 
\nabla_i Q_{ij} \nabla_k Q_{kj} + \cdots] 
\label{eq:5.3}
\en 
is the Ginzburg-Landau  
free energy  for 
${\{}Q_{ij}\}$ and  $\lambda$ is a 
space-dependent Legendre multiplier  ensuring 
 $\sum_j Q_{jj}=0$ \cite{PGbook}.   For weakly  
first order phase transition, the 
coefficient $A(T)=a_1(T-T_0) $ 
becomes small as  $T$ approaches $T_0$. 
In $F_0$ the higher order terms  
are not written  explicitly.  
Analogously to  (3.9) we find 
\be 
\delta F= \delta F_0 
-\int d{\bi r}{\bi E}\cdot\delta
{\tensor \ve}\cdot{\bi E}/8\pi.
\label{eq:5.4}\en 
Thus,  
\be 
\frac{\delta}{\delta Q_{ij}}F= 
[A(T) Q_{ij} +\lambda \delta_{ij} 
+ \cdots]  - \frac{\ve_1}{8\pi} E_iE_j,  
\label{eq:5.5}
\en 
where the gradient term is omitted. 
At  fixed capacitor potential,  
on the other hand, 
the appropriate free energy 
is  $G=F_0- 
\int d{\bi r}{\bi E}\cdot{\bi D}/8\pi$  
but  $\delta G/\delta Q_{ij}$ is of the same form 
as $\delta F/\delta Q_{ij}$
in  (5.5). In equilibrium disordered states with $T>T_0$, 
 we set $\av{\delta F/\delta Q_{ij}}=0$ to  obtain 
 \be 
\av{Q_{ij}} =   \frac{\ve_1}{8\pi} 
(E_{0i}E_{0j}-\frac{1}{3}E_0^2 \delta_{ij})\frac{1}{A(T)} 
+\cdots ,  
\label{eq:5.6}
\en 
where ${\bi E}_0$ is the average electric field assumed to be 
along the $z$ axis.  On the other hand, 
 analogously to (3.27),  
the macroscopic static dielectric constant  
is given by 
\be 
\ve_{\rm eff}= 
\av{{\ve_{zz}}} -  \int_{\bi q} 
 \frac{1}{\ve_0 q^2} \av{ 
 |\sum_j q_j {\ve}_{jz}({\small{{\bi q}}})|^2 },  
\label{eq:5.7}
\en 
where ${{\ve}_{ij}}({\small{{\bi q}}})$ are the Fourier 
transformation of ${{\ve}}_{ij}({\bi r})$. 
As $T\rightarrow T_0$, the average 
$\av{{\ve_{zz}}}$ yields  the dominant 
 contribution to the nonlinear  
dielectric constant \cite{PGbook},   
\be 
\Delta\ve_{\rm eff}\cong \ve_1^2
 E_0^2 /12\pi{A(T)}, 
\label{eq:5.8}
\en 
in agreement with the experiments \cite{liq1}. 
The  contribution from the  second term in  (5.7) 
is smaller than that in (5.8) by 
$(\ve_1/\ve_0)^2k_{\rm B}T A(T)^{1/2}/L_1^{3/2}$ 
for $L_2 \sim L_1$.

The electrostatic 
 free energy $\Delta F= F-F_0$ 
up to of order $Q^2$ is written as 
\be 
\Delta F= -\frac{E_0^2}{8\pi} \int d{\bi r} \ve_{zz}(\tensor{Q}) 
+  A_{\rm s}  \frac{E_0^2}{8\pi} \int_{\bi q} \frac{1}{q^2} |\sum_j 
{q_j}  Q_{jz}({\bi q}) |^2 , 
\label{eq:5.9}
\en  
where $A_{\rm s}=\ve_1^2/\ve_0$ and the dipolar interaction, 
the second term, is expressed in terms of the 
Fourier transformations $Q_{ij}({\bi q})$. 
The correlation functions 
of $Q_{ij}({\bi q})$ in disordered states
depend on the direction $q^{-1}{\bi q}$ 
even in the limit $q\rightarrow 0$. 
For simplicity, for $q_z=0$ we obtain 
\be 
\lim_{q\rightarrow 0} \av{Q_{iz}({\bi q})Q_{jz}({\bi q})^*} 
 = \frac{k_{\rm B}T}{r} 
 \bigg [  {\delta_{ij} } 
- \frac{q_iq_j}{q^2}   
 \frac{r_{\rm e}}{r+r_{\rm e}}  \bigg ]   \qquad (i,j=x,y), 
\label{eq:5.10}
\en 
where   
$
r= 2A(T) -  (2\ve_{21}+ \ve_{22}) E_0^2/4\pi$ 
and $ 
r_{\rm e} =  A_{\rm s} E_0^2/4\pi$.

\subsection{Director fluctuations in 
nematic states }

We consider  nematic states considerably 
below the transition,  
where  we may neglect the fluctuations of the 
amplitude $S$ \cite{PGbook}. 
Hereafter we rewrite 
$\ve_0'$  and $\ve_1'$ in (5.2) 
as $\ve_0$  and $\ve_1$ for simplicity. 
If $\ve_0$  is positive, 
 the average orientation 
of the director can be  along the $z$ axis 
from minimization of the first term in (5.8).  
Then the deviation $\delta{\bi n}$ perpendicular to 
the $z$ axis undergo large fluctuations at small 
wave numbers.   In electric field 
$\Delta F$ in (5.9) becomes  
\be 
\Delta F= \frac{1}{8\pi}E_0^2 \int_{\bi q} 
\bigg [ \ve_1 |\delta {\bi n}({\bi q})|^2 +
 A_{\rm s}'\frac{1}{q^2} |{\bi q}\cdot \delta {\bi n}({\bi q})|^2 
\bigg ] , 
\label{eq:5.11}
\en 
where $A_{\rm s}'= A_{\rm s} S^2$. For 
general ${\bi q}$ we obtain the correlation functions, 
\be 
G_{ij}({\bi q})=  \av{\delta n_i ({\bi q})\delta n_j ({\bi q})^*} 
=   k_{\rm B}T \frac{\delta_{ij} }{r_2}  
+ k_{\rm B}T \frac{q_iq_j}{q_\perp^2}   
 \bigg ( \frac{1}{r_1}-\frac{1}{r_2}  \bigg ) \quad (i,j=x,y),
\label{eq:5.12}
\en 
where   $q_\perp^2=q_x^2+q_y^2$, and 
\be 
r_1= K_3 q_z^2 + K_1 q_\perp^2 
+\frac{1}{4\pi} \bigg (\ve_1+ A_{\rm s}' \frac{q_\perp^2}{q^2} 
 \bigg ) E_0^2, \quad 
r_2= K_3 q_z^2 + K_2 q_\perp^2 + 
\frac{1}{4\pi} \ve_1E_0^2. 
\label{eq:5.13}
\en  
The $K_1$, $K_2$, and $K_3$ are the Frank constants. 
 If $\ve_1\gs A_{\rm s}'$, 
the correlation length $\xi$ is of the following order, 
\be 
\xi \sim   (4\pi K/ \ve_1)^{1/2}   E_0^{-1}, 
\label{eq:5.14}
\en 
where $K$ represents the magnitude of the Frank constants 
and $k_{\rm B}T/K$ is a microscopic length. 
The scattered light intensity is proportional to the following \cite{PGbook},  
\be 
 \av{|{\bi f}\cdot \tensor{\ve}({\small{\bi q}})\cdot{\bi i}|^2} 
=  k_{\rm B}T \ve_1^2 \bigg  [  |{\bi a}|^2/r_2  
+(1/r_1-1/r_2 ) |{\bi q}\cdot{\bi a}|^2/{q_\perp^{2}}  \bigg ] ,
\label{eq:5.15}
\en 
where ${\bi i}$ and $\bi f$ represent 
the initial and final polarizations.  The vector 
${\bi a}$ is defined by 
\be 
a_x= i_zf_x+f_zi_x, \quad a_y=i_zf_y+f_zi_y,\quad a_z=0.
\label{eq:5.16}
\en  
If  $A_{\rm s}'>0$, 
the intensity   depends on $q^{-1}{\bi q}$ even 
 in  the limit $q\rightarrow 0$.

In   nematic states 
the average  
$\av{\ve_{ij}}$ is anisotropic, 
leading to   large 
{\it intrinsic} birefringence.  We here consider 
{\it form} dichroism arising from the director fluctuations, 
where the laser wave vector $\bi k$ is along the $x$ axis and 
the average director is along the $z$ axis.  
We assume the relation 
$\ve_{ij}= \ve_0 \delta_{ij} + \ve_1 n_in_j $ 
 at optical frequencies 
using the same notation as in (5.2).  
As a generalization of  (1.5), the fluctuation contribution to the 
dielectric tensor for the electromagnetic 
waves is of the form,  
\be 
\Delta{\tensor{{\ve}}} 
= \frac{1}{\ve_0} 
\int_{\bi q} \frac{1}{q^2- k^2-i0}  \av{
 [\tensor{\ve}({\small{{\bi k}}-\small{{\bi q}}}) 
\cdot (k^2{\tensor I}- {\bi q}{\bi q}) 
\cdot \tensor{\ve}({\small{{\bi q}}-\small{{\bi k}}})]_\perp} , 
\label{eq:5.17}
\en 
where  $[\cdots]_\perp$ denotes taking the 
 tensor part perpendicular to ${\bi k}$. 
This expression reduces to (1.5) if  
$\tensor{\ve}$ is diagonal  and (1.2) holds. 
Here 
 $\Delta{{\ve}}_{yz}=0$ 
 and the imaginary part of $\Delta n /n
\cong  (\Delta{\ve_{zz}}-\Delta{\ve_{yy}})/2\ve_0$ 
becomes  
\be 
\frac{{\rm Im}\Delta n}{n} 
=\frac{k^3}{32\pi^2} (\ve_1/\ve_0)^2 
\int d\Omega \bigg [G_{xx}({\bi \ell}) 
+ {{\hat q}_z^2} G_{yy}({\bi \ell})  
- \sum_{i,j=x,y} {{\hat q}_i{\hat q}_j} G_{ij}({\bi \ell}) \bigg ],  
\label{eq:5.18}
\en 
where ${\bi \ell}  = {\bi k}-k \hat{\bi q}$ and   
$\int d\Omega \cdots $ 
denotes  integration over the direction 
$\hat{\bi q}= q^{-1}{\bi q}$. 
Using (5.12) 
we can make the following order estimations, 
\bea 
\frac{{\rm Im}\Delta n}{n} 
&\sim& 10^{-2}(\ve_1/\ve_0)^2 (k_{\rm B}T/K) 
\xi^2k^3  
\hspace{2cm} (k \xi \ll 1) \nonumber\\
&\sim& 10^{-2}(\ve_1/\ve_0)^2  (k_{\rm B}T/K)  k   
\hspace{2.6cm} (k \xi \gs  1).
\label{eq:5.19}
\ena 
The form dichroism here is much larger than 
that in (3.38) for near-critical fluids. As far as I 
am aware,  there was one attempt to measure anisotropy 
of the turbidity in  an oriented nematic state 
\cite{Langevin}.

\subsection{Orientation around a charged  particle}

As another example, we 
place a charged particle with radius $R$ and charge $Ze$ 
in a nematic state, where $\bi n$ 
is aligned along the $z$ axis or 
${\bi n} \rightarrow {\bi e}_z$  far from 
the particle. Let the density of such charged particles 
be  very low and its Coulomb potential be  not screened 
over a long distance $\lambda$ (which is the Debye 
screening length if low concentration  salt is doped). 
From (5.4) the free energy change  
 due to the orientation change ${\delta}{\bi n}$ 
is given by 
\be 
\delta F= -\int d{\bi r} \bigg [ K \nabla^2{\bi n} 
+ \frac{\ve_1}{4\pi}({\bi E}\cdot{\bi n}){\bi E} 
\bigg ]\cdot{\delta}{\bi n} ,
\label{eq:5.20}
\en 
where we assume the 
single Frank constant $K$ ($K_1=K_2=K_3=K$).  
If the coefficient $\ve_1$ is 
considerably smaller than $\ve_0$, the 
electric field ${\bi E}$ near the particle is  
of the form $-\nabla (Ze/\ve_0 r)$.  
Then, for $\ve_1>0$  (or $\ve_1<0$),  
$\bi n$ tends to be parallel 
(or perpendicular) to 
 $\hat{\bi r}=r^{-1}{\bi r}$ near the charged particle. 
We  assume that $\bi n$ is appreciably distorted from 
${\bi e}_z$ in the space region $R\ls r\ls \ell<\lambda$. 
For $\ve_1>0$ the  decrease of the 
electrostatic energy  is  estimated as   
\be 
 \Delta F \cong  -\ve_1
\frac{ Z^2e^2}{2\ve_0^2} 
\bigg (\frac{1}{R} -\frac{1}{\ell} \bigg) , 
\label{eq:5.21}
\en  
analogously to (4.4).  
For $\ve_1<0$, $\ve_1$ in (5.21) 
should be replaced by $|\ve_1|/3$ (because the 
angle  average of $({\bi e}_z\cdot{\hat{\bi r}})^2$ 
is $1/3$).  
The  Frank   free energy is estimated as 
\be 
 F_0 \sim  \pi K \ell .
\label{eq:22}
\en 
We determine $\ell$ by minimizing $F_0+ \Delta F$ to obtain 
\be
\ell = (|\ve_1|/{2\pi\ve_0^2K})^{1/2}Ze.
\label{eq:23}
\en  
The condition of strong orientation deformation 
is given by  $R<\ell$ or  
\be 
R/Z<  (|\ve_1|/{2\pi\ve_0)^{1/2} (\ell_{\rm B}}k_{\rm B}T/K)^{1/2}.
\label{eq:24}
\en 
If this  does not hold,  
the distortion of $\bi n$  
becomes weak.

\begin{figure}[t]
\vspace{-1cm}
\epsfxsize=5.0in
\centerline{\epsfbox{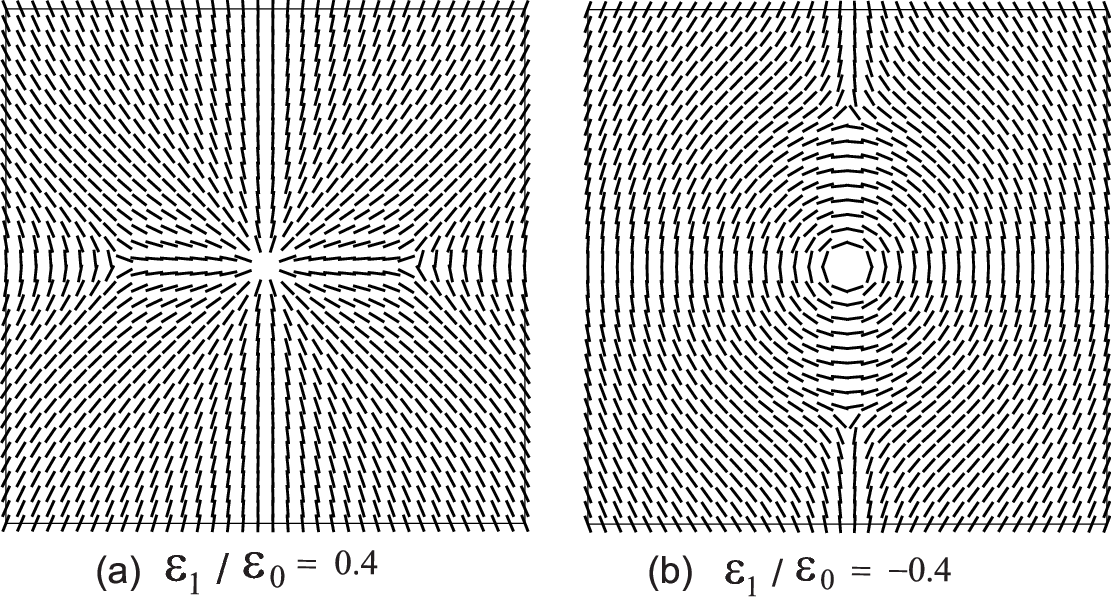}}
\caption{\protect%\narrowtext
\small{
The director field in two dimensions 
for $\ve_1/\ve_0=0.4$ 
in (a)  and 
for $\ve_1/\ve_0=-0.4$ in (b) around a charged wire 
at the origin. 
It  tends to be along  the $y$ axis (in  
the vertical direction) 
far from the origin. The spacing between 
the adjacent bars is $2\ell_{\rm B}$.   
 }}
\label{1}
\end{figure}

 Fig.6 illustrates the deformation of $\bi n$ 
in two dimensions.  
We have numerically solved     
\be 
{\bi n} \times \bigg [
K \nabla^2{\bi n} 
+ \frac{\ve_1}{4\pi}({\bi E}\cdot{\bi n}){\bi E} \bigg ] ={\bi 0}, 
\label{eq:25}
\en 
under $-\nabla \cdot\tensor{\ve}\cdot\nabla\phi=4\pi\rho$  by 
assuming ${\bi n}= (\cos\theta,\sin\theta)$ 
(or $n_z=0$). A charge  
is placed in the hard-core region $(x^2+y^2)^{1/2}<R$. 
In three dimensions this is the case of 
 an infinitely long charged wire with  radius $R$ 
and charge 
density $\sigma$, in which 
all the quantities 
depend only on $x$ and $y$. 
The solution can be characterized by 
the three normalized quantities, 
$\ve_1/\ve_0$, 
$\sigma^*\equiv  \sigma/(\ve_0K)^{1/2}$, and 
$R/\ell_{\rm B}$. Here we set  $\sigma^*=2.4$ and 
$R/\ell_{\rm B}=2$.  We discretize the space into a $200 \times 200$ 
lattice in units of $\ell_{\rm B}$ under the 
periodic boundary condition in the $x$ direction, 
so the system width is $L= 200 \ell_{\rm B}$. 
The electric potential  vanishes  at $y=0$ and $L$.

\section{Concluding remarks} 

(i) The   Ginzburg-Landau theory in Section 2 
generally  describes how the electrostatic interactions arise 
  depending on the  boundary condition (in the presence or absence 
of the capacitor plates). 
It can be  used  
to investigate   electric field effects 
at various  kinds of phase transitions in fluids and solids.  
\vspace{0.7mm}\no\\   
(ii) A  brief review  has been given on the dielectric 
properties and the electric field effects 
in near-critical fluids and liquid crystals. 
The Debye-Kleboth  experiment on 
the critical temperature shift 
was performed many years ago 
and they neglected the dipolar interaction 
in  their theoretical interpretation.   
As regards  the nonlinear dielectric constant 
$\Delta \ve_{\rm eff}$ and the birefringence 
 $\Delta n$, we cannot  explain  
their  experimental exponents,  
 $z_{\rm non} \sim  0.4$ 
and $z_{\rm op}\cong 0.65-0.8$,  
whereas  the common exponent $
 \nu(1-2\eta)\cong 0.58$ has been 
predicted for them.  To resolve these issues, 
  scattering experiments to  check 
the anisotropic structure factor (3.17) are 
 most  needed. \vspace{0.7mm}\no\\  
(iii) New predictions  have  also been made  
on  the critical temperature shift due to electric field, 
the  nonlinear dielectric constant,  
the ion effects in binary mixtures,  
and   the fluctuation intensities 
and the form dichroism  in liquid crystals. 
In particular, in 
near-critical polar mixtures with  ions, 
we have examined    
charge distributions and potential differences 
around two-phase interfaces, 
the critical temperature shift due to ions, and 
the scattering intensity.  
   The condition for  a charge-density-wave phase 
 has been examined for  general multivalent ions.  
 \vspace{0.7mm}\no\\
(iv) Effects of  oscillating electric field 
  are also worth studying particularly for   ionic systems. 
Appreciable  critical anomaly can be seen   in the 
frequency-dependence of the dielectric constant \cite{Or_this}. 
Large dynamic electric birefringence was observed  in 
polyelectrolyte solutions \cite{Hoffmann}. 
By this method we can neglect accumulation 
of ions at the boundaries (but the Joule heating may  
not be negligible).\vspace{0.7mm}\no\\  
(v) In Section 5 we have examined the deformation of 
the nematic order around a charged particle.  
The charge-induced 
orientation is intensified with decreasing 
the particle radius $R$ 
and/or increasing the charge number $Z$.
This is in marked contrast to 
 the surface anchoring of a  neutral particle \cite{Tere,exp}, 
which  can be achieved  for large radius because 
 the penalty of the Frank free energy 
needs to be small. For charged colloidal suspensions, 
furthermore,   the counterions themselves  
can  induce  large deformation 
of the nematic order  (because of their small size) and  
tend to accumulate near the large particles. 
These aspects should be examined in future. 
\vspace{0.7mm}\no\\ 
(vi) Stronger electric field effects have been observed 
in polymeric systems  
than in low-molecular-weight 
fluids. As such effects,  
we mention field-induced anisotropy in  light scattering 
from  a polymer solution \cite{Fuller_e1},
 lamellar alignment 
in diblock copolymers \cite{Helfand}, and 
large dielectric response in a surfactant sponge phase \cite{Cates}.

\vspace{5mm}
 
\no {\Large{\bf Acknowledgments}}\\

\vspace{-2mm}
\no I would like to thank 
S.J.  Rzoska  for valuable discussions 
on the electric field effects in fluids. 
Thanks are also due to 
K. Orzechowski for guidance of the 
ion effects in binary mixtures and for showing 
 Refs.$[41$-$43]$. 
G.G. Fuller informed me of  Ref.\cite{Langevin}.  

\end{document}